\documentclass[english,11pt]{article}
\pdfoutput=1
\usepackage{geometry}
\usepackage[sort,comma]{natbib}
\usepackage[pagebackref=false,colorlinks]{hyperref}
\hypersetup{pdffitwindow=true,linkcolor=blue,citecolor=blue,urlcolor=blue}
\usepackage{amsmath}
\usepackage{bbm}
\usepackage{multirow}
\usepackage{graphicx}
\DeclareGraphicsExtensions{.pdf,.jpg,.png}
\usepackage{subfigure}

\usepackage{amsfonts}
\usepackage{mathdots}
\usepackage{amsthm}
\usepackage{newtxmath}

\usepackage{multirow}
\usepackage{centernot}
\usepackage{todonotes}
\usepackage[ruled,linesnumbered]{algorithm2e}
\usepackage{float}

\SetKwInOut{Input}{Input}
\SetKwInOut{Output}{Output\,}
\SetKwInOut{Data}{Data}

\DeclareMathOperator*{\argmax}{arg\,max}

\usepackage[font={footnotesize,it}, sc, bf]{caption}
\renewenvironment{abstract}{%
  \noindent\textbf\abstractname .\hspace{1pt}
}{
  \endlist \par\bigskip\bigskip
}

\usepackage{lastpage}
\RequirePackage[hyperpageref]{backref}
\renewcommand*{\backref}[1]{} 
\renewcommand*{\backrefalt}[4]{
    \ifcase #1
       No referred.
    \or
       \emph{Referred to on page #2.}
    \else
       \emph{Referred to on pages #2.}
    \fi}

\usepackage[shortlabels]{enumitem}

\usepackage[titletoc,title]{appendix}

\binoppenalty=\maxdimen
\relpenalty=\maxdimen

\begin{document}

\begin{center}
{\LARGE\bf Fast approximate inference for variable selection in Dirichlet process mixtures, with an application to pan-cancer proteomics}
\end{center}
\medskip
\begin{center}
{\large Oliver M. Crook$^{1,2,3}$, Laurent Gatto$^{4}$ and Paul D. W. Kirk$^{3}$ \\[15pt]
\emph{$^{1}$Department of Applied Mathematics and Theoretical Physics, University of Cambridge, U.K.}\\
\emph{$^{2}$Cambridge Centre for Proteomics, Department of Biochemistry, University of Cambridge, U.K.}\\
\emph{$^{3}$MRC Biostatistics Unit, School of Clinical Medicine, University of Cambridge, U.K.}\\
\emph{$^{4}$de Duve Institute, UCLouvain, Belgium}\\
}
\end{center}

\bigskip

\begin{center}
Preprint, \today
\end{center}
\bigskip\bigskip

\begin{abstract}
The Dirichlet Process (DP) mixture model has become a popular choice for model-based clustering, largely because it allows the number of clusters to be inferred.  The sequential updating and greedy search (SUGS) algorithm \citep{Wang::2011} was proposed as a fast method for performing approximate Bayesian inference in DP mixture models, by posing clustering as a Bayesian model selection (BMS) problem and avoiding the use of computationally costly Markov chain Monte Carlo methods.  Here we consider how this approach may be extended to permit variable selection for clustering, and also demonstrate the benefits of Bayesian model averaging (BMA) in place of BMS. Through an array of simulation examples and well-studied examples from cancer transcriptomics, we show that our method performs competitively with the current state-of-the-art, while also offering computational benefits.  We apply our approach to reverse-phase protein array (RPPA) data from The Cancer Genome Atlas (TCGA) in order to perform a pan-cancer proteomic characterisation of 5,157 tumour samples.   We have implemented our approach, together with the original SUGS algorithm, in an open-source R package named sugsvarsel, which accelerates analysis by performing intensive computations in C++ and provides automated parallel processing.  The R package is freely available from: \url{https://github.com/ococrook/sugsvarsel}
\end{abstract}
%\null\bigskip
%\tableofcontents

%%%%%%%%%%%%%%%%%%%%%%%%%%%%%%%%%%%%%%%%%%%%%%%%
%%%%%%%%%%%%%%%%%%%%%%%%%%%%%%%%%%%%%%%%%%%%%%%%

\section{Introduction}\label{sec:intro}
Bayesian nonparametric methods have become commonplace in the statistics and machine learning literature due to their flexibility and wide applicability. For model-based clustering, Dirichlet process \citep{ferguson::1973,ferguson::1974} mixture models have become particularly popular \citep{antoniak::1974, lo::1984, Escobar::1994, West::1995, Blei::2006}, %. These methods place a Dirichlet process (DP) \citep{ferguson::1973,ferguson::1974} prior over parameters in a mixture model, 
partly because they allow the number of clusters supported by the data to be inferred.  By introducing latent selection indicators, these models can be extended to perform variable selection for clustering \citep{Kim::2006}, which is particularly relevant in high-dimensional settings %, finding relevant clusterings in the data is made challenging by the presence of nuisance or irrelevant variables, which can not only degrade the quality of the clusters but make them difficult to interpret
 \citep{Law::2004,Const::2006}. There are now several approaches for model-based clustering and variable selection \citep[see][for a recent review]{fop::2017}, but current Markov chain Monte Carlo (MCMC) algorithms for Bayesian inference in Dirichlet process (DP) mixture models \citep[e.g.][]{Neal::2000, Jain::2004} are computationally costly, and often infeasible for large datasets.
\\
\\
Algorithms for fast approximate inference in DP mixture models, such as the use of fast search algorithms \citep{Daume::2007}, Bayesian hierarchical clustering \citep{Heller::2005,Savage::2009,Cooke::2011,Darkins::2013}, or the sequential updating and greedy search (SUGS) algorithm \citep{Wang::2011,vsugs::2014}, make possible the analysis of datasets with large numbers of observations. However, without variable selection such algorithms may be ill-suited to the high-dimensional setting. In the spirit of the original SUGS algorithm, here we pose clustering and variable selection as a Bayesian model selection (BMS) problem. We consider variable selection for clustering in terms of partitioning variables into those which are relevant and those which are irrelevant for defining the clustering structure, and thereby pose the problem as one of using BMS to select both a partition of the variables and a partition of the observations.  We moreover consider the benefits of performing Bayesian model averaging (BMA) \citep{Madigan::1994, Hoeting::1999} for summarising the SUGS output.  For ease of exposition, we focus on the case of DP Gaussian mixtures, but note that all of our methods extend straightforwardly to other distributions for which conjugate priors may be chosen.
\\
\\
We consider a range of simulation settings and well-studied examples from cancer transcriptomics to show that our methods perform competitively with the current state-of-the-art.  Having established the utility of our approach, we consider an application to reverse-phase protein arrays (RPPA) datasets in order to characterise the pan-cancer functional proteome.  Such datasets have the potential to provide a deeper understanding of the biomolecular processes at work in cancer cells, and have previously been shown to offer additional insights beyond what may be captured by genomics or transcriptomics datasets \citep{Akbani::2014}.  Here we consider RPPA data for 5,157 tumour samples obtained from The Cancer Genome Atlas (TCGA). 
\\
\\
Section \ref{sec:meth} recaps DP mixture models and the SUGS algorithm, then describes our extensions to SUGS including variable selection and BMA. Section \ref{sec:verify} evaluates our method on simulated datasets and compares it with other approaches to clustering and variable selection.  We then apply our method to a large proteomics dataset, highlighting its applicability. In the final section, we make some concluding remarks and discuss limitations and extensions. Our methods are implemented in an R package: \url{https://github.com/ococrook/sugsvarsel}.

%%%%%%%%METHODS:
\section{Methods}
\label{sec:meth}
\subsection{Dirichlet process mixtures}
We provide a very brief recap of DP mixture models, mainly to introduce notation, and refer to the overview provided in Section 3 of \citet{Teh:2005} for further details. Let $G \sim DP(\beta P_0)$ where $\beta>0$ is the DP concentration parameter, $P_0$ is the base probability measure, and $G$ is a random probability measure.  We consider a P\'olya urn scheme in which we have independent and identically distributed (i.i.d.) random variables $\theta_1,\theta_2,...$ distributed according to $G$. Computing the sequential conditional distributions of $\theta_i$ given $\theta_1,...,\theta_{i-1}$, upon marginalising out the random $G$, we obtain \citep{blackwell::1973}:
\begin{equation}\label{equation::Polya}
\theta_i|\theta_1,...,\theta_{i-1} \sim \frac{\beta}{\beta+i-1}P_0 + \frac{1}{\beta+i-1}\sum_{l=1}^{i-1}\delta_{\theta_l},\,\,i=1,...,n,
\end{equation}
where $\delta_{\theta}$ is a probability measure with mass concentrated at $\theta$.  It is clear from this equation that for any $r = 1, 2, \ldots, i-1$, the probability that $\theta_i$ is equal to $\theta_r$ is given by ${\sum_{l = 1}^{i-1} \mathbb{I}(\theta_l = \theta_r)}/({\beta+i-1})$, where $\mathbb{I}(X) = 1$ if $X$ is true and $\mathbb{I}(X) =0$ otherwise.  Thus $\theta_i$ has non-zero probability to be equal to one of the previous draws, and it is this clustering property that makes the DP a suitable prior for mixture models.     

The DP mixture model is obtained by introducing an additional parametric probability distribution, $F$. More precisely, let observations $x_i$ be modelled according to the following hierarchical model:
\begin{equation}
\begin{split}
G &\sim DP(\beta P_0),\\
\theta_i|G &\sim G,\\
x_i|\theta_i &\sim F(\theta_i),
\end{split}
\end{equation}
where $F$ denotes the conditional distribution of the observation $x_i$ given $\theta_i$. For example, when $F$ is chosen to be a Gaussian distribution we arrive at the DP Gaussian mixture model \citep[also referred to as the infinite Gaussian mixture model;][]{rasmussen::2000}. %The key assumption is that there is an infinite sequence of components, with $\theta_k$ denoting the parameters for component $k$, for $k=1,...,\infty$. Observations can be sequentially allocated to a finite realisation of these components using Equation~\eqref{equation::Polya}, with the first observation $x_1$ being arbitrarily assigned to cluster $k=1$. 

When performing inference for such models, it is common to introduce a set of latent variables (cluster labels) $z_1,...,z_n$ associated with the observations, such that $z_i$ is the cluster label for observation $x_i$. From the above specification of the DP mixture model, it follows that the conditional prior distribution of $z_i$ given $z_{-i} = (z_1,...,z_{i-1})$ is categorical with:
\begin{equation}\label{equation::dirichletpredict}
\pi_{ik}:=P(z_i = k|z_{-i}, \beta) = \begin{cases}\frac{n_k}{\beta+i-1}, & \mbox{for } k=1,..,K-1 \\ \frac{\beta}{\beta+i-1}, & \mbox{for } k=K, \end{cases}
\end{equation}
where $\beta>0$ is the DP concentration parameter, $n_k := {\sum_{l = 1}^{i-1} \mathbb{I}(z_l = k)}$ is the number of previous observations allocated to cluster $k$,  and $K=\max\{z_{-i}\}+1$. Larger values of $\beta$ encourage observations to be allocated to new clusters, hence $\beta$ plays a role in controlling the number of clusters. 

%To complete the specification of the model, we place priors on the parameters for each cluster:
%\begin{equation}
%p(\theta)=\prod_{k=1}^{\infty}p_0(\theta_k),
%\end{equation}
%where $p_0$ is the prior distribution on the cluster specific parameter $\theta_k$. 

Inference for DP mixture models can performed using computationally intensive MCMC methods \citep{Neal::2000, Jain::2004}.  However, as we discuss, here we are interested in the SUGS algorithm for approximate inference, proposed by \cite{Wang::2011}.
\subsection{Sequential Updating and Greedy Search (SUGS)}
SUGS is a sequential approach for allocating observations to clusters, which (greedily) allocates the $i$-th observation to a cluster, given the allocations of the previous $i-1$ observations.  Suppose that observations $x_{-i} = (x_1,...,x_{i-1})$ have previously been allocated to clusters.  As described in \citet{Wang::2011}, the posterior probability of allocating observation $i$ to cluster $k$ according to the DP mixture model formulation above is given by:
\begin{equation}\label{equation::posteriorallocat}
P(z_i=k|x_i,x_{-i},z_{-i}, \beta) = \frac{\pi_{ik}L_{ik}(x_i)}{\sum_{l=1}^{K}\pi_{ik}L_{il}(x_i)},
\end{equation}
where $\pi_{ik}$ is defined as in Equation \eqref{equation::dirichletpredict}, and 
\begin{equation}
L_{ik}=\int f(x_i|\theta_k)p(\theta_k|x_{-i},z_{-i})\,d\theta_k\label{Lik}
\end{equation}
is the conditional marginal likelihood associated with $x_i$ given allocation to cluster~$k$ and the cluster allocations for observations $1,...,i-1$, with $f(x_i|\theta_k)$ denoting the likelihood associated with $x_i$ as a function of $\theta_k$. If $k$ is a cluster to which previous observations have already been allocated, then $p(\theta_k|x_{-i},z_{-i}) $ is the posterior distribution of $\theta_k$ given the observations previously allocated to cluster $k$; i.e. $p(\theta_k|x_{-i},z_{-i}) \propto p_0(\theta_k)\prod_{j:z_j=k, 1\leq j\leq i-1}f(x_j|\theta_k)$, where $p_0(\theta_k)$ is the prior on the cluster-specific parameters, $\theta_k$. For a new cluster, i.e. for $k=K$, we have $p(\theta_k|x_{-i},z_{-i})=p_0(\theta_k)$. If $p_0$ is taken to be conjugate for the likelihood $f$, then the posterior and conditional marginal likelihood are available analytically.

Assuming that the concentration parameter $\beta$ is given and that conjugate priors are taken, the above suggests a computationally efficient deterministic clustering algorithm (the SUGS algorithm).  That is, $z_1$ is initialised as $z_1 =1$, and then subsequent observations are sequentially allocated to clusters by setting $z_i = \argmax_{k \in \{1, \ldots, K \}} P(z_i=k|x_i,x_{-i},z_{-i}, \beta)$, where we recall that $K=\max\{z_{-i}\}+1$ may change after each sequential allocation. 
 
\subsubsection{Dealing with unknown $\beta$ }
The DP concentration parameter $\beta$ directly influences the number of clusters, thus we treat this as a random variable to be inferred, in the same way as in \cite{Wang::2011}.  In particular, let $\boldsymbol{\hat{\beta}} = (\hat{\beta}_1,...,\hat{\beta}_L)$ be a discrete grid of permissible values for $\beta$ with a large range, and then define the prior for $\beta$ to be discrete with the following form:
\begin{equation}
p_0(\beta | \kappa_1, \ldots, \kappa_L) = \sum_{l = 1}^{L} \kappa_l \mathbb{I}(\beta = \hat{\beta}_l),
\end{equation} 
where $\kappa_l = p(\beta = \hat{\beta}_l)$. Further defining $\phi_l^{(i-1)}  =   p(\beta = \hat{\beta}_l| x_{-i}, z_{-i})$ and $\pi_{ikl} = p(z_i = k| \beta = \hat{\beta}_l , z_{-i})$, the $\beta$ parameter may be marginalised in Equation \eqref{equation::posteriorallocat} to obtain:
\begin{equation}\label{equation::polyarandombeta}
p(z_i = k |x_{-i}, x_i, z_{-i}) = \frac{\sum_{l = 1}^{L}\phi_l^{(i-1)} \pi_{ikl} L_{ik}(x_i)}{\sum_{l = 1}^{L}\phi_l^{(i-1)} \sum_{k=1}^{K} \pi_{ikl} L_{ik}(x_i)}, 
\end{equation}
where $\pi_{ikl} := p(z_i = k| \beta = \hat{\beta}_l , z_{-i})$ is given by Equation \eqref{equation::dirichletpredict}; $\phi_l^{(0)} = \kappa_l$; and:
\begin{equation}\label{equation::sugsbetaudpates}
\phi_l^{(i)} =   p(\beta = \hat{\beta}_l| x_{-i},x_i,z_{-i},z_i) =\frac{\phi_l^{(i-1)}\pi_{iz_il}}{\sum_{s=1}^{L}\phi_s^{(i-1)}\pi_{iz_is}}
\end{equation}
may be calculated sequentially for $i = 1, \ldots, n$.  The SUGS algorithm for allocating observations to clusters when $\beta$ is unknown is then as presented in Algorithm \ref{alg1}.
\newline

\begin{algorithm}[H]
  \caption{The SUGS algorithm, when the DP precision parameter $\beta$ is allowed to be unknown.}\label{alg1}
  \Input{Data $X = \{x_i\}_{i = 1}^n$, Prior $p_0(\theta),$\newline Hyperparameters $\{\kappa_l\}_{l=1}^L$}
  \Output{Cluster allocations $Z = \{z_i \}_{i = 1}^n$}
  Initialise $z_1 = 1$, $K = 2$, and $\{\phi^{(0)}_l=\kappa_l\}_{l=1}^L$;
  
Evaluate $p(\theta_1 | z_1, x_1) \propto p_0(\theta_1)f(x_1|\theta_1)$;

Calculate $\{\phi^{(1)}_l\}_{l=1}^L$, according to Eq. \eqref{equation::sugsbetaudpates};

\For {$i = 2$ to $N$} {
\For {$k = 1$ to $K$} {
		Calculate $L_{ik}$ according to Eq. \eqref{Lik};

		Evaluate $p(z_i=k|x_{1}, \ldots,x_i,z_1, \ldots, z_{i-1})$ according to Eq. \eqref{equation::polyarandombeta};
		}
	
		Set $z_i = \argmax_{k= 1,\ldots, K} (p(z_i=k|x_{1}, \ldots,x_i,z_1, \ldots, z_{i-1})) $;

		Set $K=\max\{z_{1}, \ldots, z_{i}\}+1$;

\For {$l = 1$ to $L$} {
		Calculate $\phi_l^{(i)}$, according to Eq. \eqref{equation::sugsbetaudpates};
		}

		Evaluate $p(\theta_{z_i}|x_{1}, \ldots, x_i,z_{1}, \ldots z_i) \propto p_0(\theta_{z_i})\prod_{j:z_j=z_i, 1\leq j\leq i}f(x_j|\theta_{z_i})$;		
		
    }
\end{algorithm}

\subsubsection{Formulation of Bayesian Model Selection problem}\label{BMSsection}
A notable limitation of the (deterministic) SUGS algorithm as presented so far is that the clustering structure obtained is dependent upon the initial ordering of the observations. To remove this dependence, \cite{Wang::2011} consider multiple permutations of this ordering, and pose SUGS as a Bayesian model selection (BMS) problem. More concretely, the algorithm is repeated for many random orderings of the data and a final partition of the observations is then chosen by optimising an appropriate objective function for BMS, such as the marginal likelihood (ML):
\begin{equation}
L(X|Z) = \prod_{k=1}^{K}\int_{\theta_k}\left[\prod_{i:z_i = k} f(x_i|\theta_k)\right]p_0(\theta_k)d\theta_k. 
\end{equation}
In practice, \citet{Wang::2011} advocate optimising the {\em pseudo}-marginal likelihood (PML), since they found that the marginal likelihood to often produce many small clusters.  The PML is given by:
\begin{equation}\label{equation::PML}
\begin{split}
\mbox{PML}_z(X) = & \prod_{i=1}^{N}p(x_i|X_{n\backslash-i},z_{n\backslash-i} ) \\
=& \prod_{i=1}^{N}\int_{\theta}p(x_i|\theta)p(\theta|X_{n\backslash-i},z_{n\backslash-i}) d\theta \\
= & \prod_{i=1}^{N}\sum_{k=1}^{K}P(z_i=k|X_{n\backslash-i},z_{n\backslash-i})\int_{\theta_k} f(x_i|\theta_k)p(\theta_k|X_{n\backslash-i},z_{n\backslash-i})d\theta_k,
\end{split}
\end{equation}
where, defining $X = \{x_1,...,x_n\} $ and $Z = \{z_1,...,z_n\}$, we have $X_{n\backslash-i} = X\backslash \{x_i\}$ is the set of all observations except the $i^{th}$, and similarly $z_{n\backslash-i} =  Z\backslash \{z_i\} $.   In addition, \cite{Wang::2011} remark that that $p(x_i| X, Z)$ can be used to approximate $p(x_i|X_{n\backslash-i},z_{n\backslash-i} )$ to speed up computations and that this approximation is accurate for large sample sizes.
 \subsection{SUGS for variable selection}
 Irrelevant variables in high-dimensions can present a considerable challenge for clustering models and algorithms, because the number of variables with no clustering structure can overwhelm those where a clustering structure exists \citep{Witten:2010}. There have been many approaches to model-based clustering and variable selection \citep[e.g.][]{Raftery::2006, Maugis::2009}, and we direct readers to \cite{fop::2017} for a recent review. However, many of these scale poorly with increasing dataset dimension, and/or require the number of clusters to be determined as a separate analysis step.  To address these challenges, here we extend the SUGS algorithm to simultaneously perform clustering and variable selection, and refer to the resulting procedure as {\em SUGSVarSel}. 
  %The advantage of a Bayesian nonparametric approach is we do not need to specify the number of clusters beforehand or even the DP concentration parameter. Currently methods rely on computationally intensive MCMC algorithms and as the number of variables and observations increase they become infeasible to apply.
 \\
 \\
Since we are in the high-dimensional setting, we assume for simplicity that variables are independent given the cluster allocations (which, in the Gaussian case, is equivalent to assuming a diagonal structure for the covariance matrix). Let $x_{i,d}$ be the $d^{th}$ element of the $i^{th}$ observation vector, with $d = 1, \ldots, D$, and $D$ the number of variables. Introducing indicator variables $\gamma_d$, which is $1$ if the $d^{th}$ variable is relevant for the clustering structure and $0$ if not, we follow a common approach from the literature \citep{Law::2004, Tadesse::2005, Kim::2006} and assume that the cluster conditional likelihood can be factorised as follows:
\begin{equation}
f(x_i|\theta, \theta_{0},z_i= k) = \prod_{d=1}^{D}f(x_{i,d}|\theta_{k,d})^{\mathbb{I}(\gamma_d = 1)}f(x_{i,d}|\theta_{0,d})^{\mathbb{I}(\gamma_d = 0)},\label{clustCondLike}
\end{equation}
where $\theta_0$ are ``global'' (i.e. not cluster-specific) parameters.  In other words, the variables for which $\gamma_d = 1$ are modelled by a mixture distribution with cluster-specific parameters $\theta_{k,d}$, while the variables for which $\gamma_d = 0$ are modelled by a single component with (global, not cluster-specific) parameters $\theta_{0,d}$.  Having introduced the $D$ indicator variables $\gamma_d$, we now extend the SUGS algorithm in order to estimate them.   
\subsubsection{The SUGSVarSel algorithm}
Given a realisation of the indicator variables, $\Gamma = \{\gamma_1, \ldots, \gamma_D\}$, we may plug the cluster conditional likelihood given in Equation \eqref{clustCondLike} into Equation \eqref{Lik} and proceed as before in order to identify a clustering, $Z$.  

Conversely, suppose we have a realisation, $Z$, of the set of component allocation variables, but that the indicator variables $\Gamma$ are unknown.  In this case, the posterior probabilities associated with the variable indicators are given by:
\begin{align} 
P(\gamma_d =1|X,Z) &= \frac{p_0(\gamma_d =1)}{B}\prod_{k\in Z}\, \int_{\theta_{k,d}}\left(\prod_{i:z_i = k} f(x_{i,d}|\theta_{k,d})\right)p_0(\theta_{k,d})d\theta_{k,d}\label{equation::SUGSfeaton}\\
P(\gamma_d = 0|X,Z) &= \frac{p_0(\gamma_d = 0)}{B} \int_{\theta_{0,d}}\left(\prod_{i:z_i = k} f(X_{d}|\theta_{0,d})\right)p_0(\theta_{0,d})d\theta_{0,d}\label{equation::SUGSfeatoff},
\end{align}
where $p_0(\gamma_d = q)$ indicates the prior probability that $\gamma_d = q$, and $B$ is a normalising constant that ensures that $p(\gamma_d =0|X,Z)$ and $p(\gamma_d =1|X,Z)$ sum to 1.  Thus, given a realisation, $Z$, of the set of component allocation variables, a greedy approach to finding $\gamma_d$ is to set $\gamma_d = \argmax_{q \in \{0, 1 \}} P(\gamma_d = q|X,Z)$.

Given an initial realisation of the indicator variables, $\Gamma = \Gamma^{(0)}$,  the above suggests an iterative strategy in which at each iteration we use the SUGS algorithm to find a partition $Z^{(t)}$ given $\Gamma^{(t-1)}$, and then greedily update the indicator variables according to Equations \eqref{equation::SUGSfeaton} and \eqref{equation::SUGSfeatoff} above in order to obtain $\Gamma^{(t)}$ given $Z^{(t)}$.  This algorithm, which we refer to as SUGSVarSel, is presented in Algorithm \ref{alg2}.
\\
%Our extension to the SUGS algorithm rests on the following key ideas. First fix an ordering of the observations and produce a clustering with the SUGS algorithm. Then partition variables into those that support a clustering structure and those that support the null model using equations (\ref{equation::SUGSfeaton}) and (\ref{equation::SUGSfeatoff}). We can then iterate between clustering and variable selection repeatedly until there are no changes between iterations. Thus, our version of the SUGS algorithm which we call SUGS with variable selection (SUGSVarSel) takes the following form.

\begin{algorithm}[H]
  \caption{The SUGSVarSel algorithm}\label{alg2}
  \Input{Data $X = \{x_i\}_{i = 1}^n$, Priors $p_0(\theta)$ and $p_0(\gamma)$, \newline Hyperparameters $\{\kappa_l\}_{l=1}^L$, Initial Indicator Switches $\Gamma^{(0)}$, Maximum Iterations $T$.}
  \Output{Cluster allocation $Z = \{z_i \}_{i = 1}^n$ \newline 
  	 	Variable switches $\Gamma = \{\gamma_d \}_{d = 1}^D$}
  Initialise $z_1 = 1$, $K = 2$, and $\{\phi^{(0)}_l=\kappa_l\}_{l=1}^L$;
  
Evaluate $p(\theta_1 | z_1, x_1) \propto p_0(\theta_1)f(x_1|\theta_1)$;

Calculate $\{\phi^{(1)}_l\}_{l=1}^L$, according to Eq. \eqref{equation::sugsbetaudpates};

\While{$t \leq T$}{
    \For {$i = 2$ to $N$} {
    	\For {$k = 1$ to $K$} {
    	Calculate $L_{ik}$ given $\Gamma^{(t - 1)}$, according to Eqs. \eqref{Lik} and \eqref{clustCondLike};
    	
   		 Evaluate $p(z_i=k|x_{1}, \ldots,x_i,z_1, \ldots, z_{i-1})$ according to Eq. \eqref{equation::polyarandombeta};
   	} 
    	
			Set $z_i = \argmax_{k= 1,\ldots, K} (p(z_i=k|x_{1}, \ldots,x_i,z_1, \ldots, z_{i-1})) $;
		
			Set $K=\max\{z_{1}, \ldots, z_{i}\}+1$;

		\For {$l = 1$ to $L$} {
			Calculate $\phi_l^{(i)}$, according to Eq. \eqref{equation::sugsbetaudpates};
		}
			Evaluate, using the cluster conditional likelihood in Eq. \eqref{clustCondLike}, $p(\theta_{z_i}|x_{1}, \ldots, x_i,z_{1}, \ldots z_i) \propto p_0(\theta_{z_i})\prod_{j:z_j=z_i, 1\leq j\leq i}f(x_j|\theta_{z_i})$;
		
	}
    
    \For {$d = 1$ to $D$} {
    	Calculate $p(\gamma_d=r|X,Z)$, according to Eqs. \eqref{equation::SUGSfeaton} and \eqref{equation::SUGSfeatoff}; 
		
		Set $\gamma_{d}= \argmax_{ r \in \{0,1\}}(p(\gamma_d=r|X,Z))$;
		
    }
    $t \gets t+1$
}    
\end{algorithm}

%\begin{algorithm}[H]
%\begin{enumerate}
%	\item Input: Data $X$, initial $\Gamma^{(0)}$, maximum iterations $T$
%	\item Set $\phi^{(0)}_l=\kappa_l$ and set $z_1 = 1$
%	\item Calculate $p(\theta_1 | z_1, x_1) \propto p_0(\theta_1)f(x_1|\theta_1)$
%	\item Update $\phi_l^{(1)}$
%	\item While $t \leq T$
%	\begin{enumerate}
%	
%	\item For $i = 2,...,N$
%	\begin{itemize}
%		\item choose $z_i =k$, where $k = \max_k (p(z_i=k|x_{-i},x_i,z_{-i})) $
%		\item update $p(\theta_{z_i}|x_{-i},z_{-i})$ by using the data $x_i$ and $z_i$
%		\item  update $\phi_l^{(i)}$
%	\end{itemize}
%	
%	\item  For $d = 1,...,D$
%	\begin{itemize}
%		\item Assign $\gamma_{d}=r$, where $r=\max_{\{0,1\}}(p(\gamma_d=r|X,Z))$
%	\end{itemize}
%	\item Set $t = t+1$ 	 
%	\end{enumerate}
%	\item Return $Z = \{z_i \}_{i = 1}^n$ and $\Gamma = \{\gamma_d \}_{d = 1}^D$.
%\end{enumerate}
%\caption{The SUGSVarSel algorithm.}\label{alg2}
%\end{algorithm}
\subsubsection{Initialisation strategies for SUGSVarSel}
%The clustering and variable partitions obtained are interdependent. Randomising the ordering of the data allows us to visit many possible partitions of the observations. However, the quality of the clustering obtained also depends on the initial partitioning of the variables. %We note that in applications an ``all-on'' initialisation, that is setting $\gamma_d=1$, for all $d=1,..,D$, can lead to lack of exploration of many models and thus leads to a poor clustering and variable selection for the data. 
Like the SUGS algorithm, the output of SUGSVarSel depends upon the initial ordering of the observations.  It moreover depends upon the initialisation of the variable selection switches, $\Gamma^{(0)}$.  To address this latter issue, we propose a random sub-sampling initialisation strategy. This is as follows: first randomly select $p_1$ variables (with $1 < p_1 \le D$) and apply SUGSVarSel on this new dataset $\tilde{X}$ of size $n \times p_1$ with a small number of random orderings of the observations (we find $10$ works in practice). The initial indicator for the variables of $\tilde{X}$, which we write as $\tilde{\Gamma}^{(0)}$, are set as all-on ($\gamma_{d} = 1 $ for these $p_1$ variables). $\tilde{\Gamma}^{(0)}$ is held the same for each of the random orderings. For each of the random orderings, this approach outputs $\tilde{Z}$ for all observations but $\tilde{\Gamma}$ for only a subset of size $p_1$ of the variables. To obtain $\Gamma$ for all $D$ variables, we use the cluster allocations $\tilde{Z}$ and the full data $X$ to compute probabilities for the remaining variables using equations \ref{equation::SUGSfeaton} and \ref{equation::SUGSfeatoff}. We then greedily assign the indicator variables. A single best model generated by these random orderings is selected using the ML. This procedure returns a $\Gamma_1 \in \{0,1\}^{D}$; that is, variable selection switches with some variables switched on and other variables switched off.
%{\color{red} [What was $\Gamma^{(0)}$ for each of these 10 runs?  Does ``first randomly select $p_1$ variables'' mean that $\Gamma^{(0)}$ is obtained by setting $\gamma_d = 1$ for these $p_1$ variables?  Or does it mean that we run SUGSVarSel on a reduced problem in which we only keep $p_1$ variables from the original dataset?  If the latter case, what is $\Gamma^{(0)}$?  Do we have the same $\Gamma_0$ for each of the 10 random orderings, or a different one each time?]}
%
We repeat this process for a total of $M$ random sub-samples of the variables to produce a set of clusterings ${Z}_1,...,{Z}_M$ and a set of variables $\Gamma_1,..,\Gamma_M$. These variable sets are then used as initial inputs $\Gamma^{(0)} = \Gamma_i$ for $i = 1,...,M$ for the SUGSVarSel algorithm (which is now run using all variables $p = D$) with $Q$ new random orderings (again we find $10$ is sufficient in practice). This SUGSVarSel with sub-sampling initialisation strategy returns $Q$ models for each random sub-sample of the variables. Thus, we have $QM$ models from which to choose.
%
%{\color{red} [This is the bit I find most confusing, because I feel like using $\Gamma_i$ as an initial input means that some variables are initially switched on and some are initially switched off, but then saying that the SUGSVarSel algorithm is run using all variables ($p = D$) could be understood to mean that {\em all} the variables are switched on and {\em none} are switched off.  Also, what are the orderings of the data when SUGSVarSel is run with $\Gamma_i$ as the input?  Are multiple orderings considered for each $\Gamma_i$, and, if so, how many?]}
%{\color{red} [It might help to spell out how many models we have at this stage.  Is it $M$, or more than $M$?  If $Q$ random orderings are considered for each $\Gamma_i$, does that mean that we have $MQ$ models at this stage?]}
%
For each model obtained in this way, we calculate the marginal likelihood (see Section~\ref{BMSsection}).  We can then perform BMS to obtain a single ``best'' model, or we can use Bayesian model averaging (BMA; see next section).

\subsection{Bayesian Model-Averaged Co-clustering Matrices}
\subsubsection{Bayesian model averaging}
The output of our algorithm is a set of clusterings, associated variables and a marginal likelihood. One can select a single ``best'' model amongst these possible clustering, however we can also average over these models to capture the model uncertainty. The idea is called Bayesian model averaging (BMA) and we apply the method to clustering and variable selection \citep{Madigan::1994, Hoeting::1999,Russell::2015}. %BMA is not directly applicable in our setting, since the labelling of clusters can arbitrarily change between models - the so-called label switching problem \citep{Jasra::2005, Richardson::1997, Stephens::2000}. Furthermore, in our case, the number of clusters between models need not be the same. Thus BMA is only indirectly applicable; that is, we average so-called co-clustering matrices rather than directly averaging the models. 

For each model we form a co-clustering matrix $S$. $S$ is defined in the following way:
\begin{equation}
S_{ij}= \begin{cases}0, & \mbox{if } z_i\neq z_j \\ 1, & \mbox{if } z_i=z_j. \end{cases}
\end{equation}
That is the $ij^{th}$ entry of $S$ is $1$ if observation $x_i$ and $x_j$ are in the same cluster and $0$ otherwise. We note that the $S$ is invariant to relabelling and the number of clusters. Now, suppose we have $M$ models $\mathcal{M}_1,...,\mathcal{M}_M$, letting $X$ be our observations and $\theta_m$ be the parameters associated with model $\mathcal{M}_m$. The posterior probability for $\mathcal{M}_m$ is given by
\begin{equation}\label{equation::BMA}
p(\mathcal{M}_m|X) = \frac{p(X|\mathcal{M}_m)p_0(\mathcal{M})}{\sum_{l=1}^{M}p(X|\mathcal{M}_l)p_0(\mathcal{M}_l)},
\end{equation}
where
\begin{equation}\label{equation::BMAML}
P(X|\mathcal{M}_m) = \int P(X|\theta_m, \mathcal{M}_m)P(\theta_m|\mathcal{M}_m)\,d\theta_m.
\end{equation}
The marginal likelihood (\ref{equation::BMAML}) is the key quantity for model comparison and can be interpreted as the weight given to each proposed model. Further note the two sources of averaging: the averaging over the parameters in the ML and the averaging over the models in equation (\ref{equation::BMA}). We suppose that \textit{a priori} all models are equally likely, choosing the prior on each model to be $p_0(\mathcal{M}_m)=1/M$. One computational challenge that (\ref{equation::BMA}) gives us is computing the summation, since it can involve evaluating possibly thousands of models. To overcome this, one can discount models that are poor at describing our observations comparatively to our best model. More precisely, let us form Occam's window \citep{Hoeting::1999}:
\begin{equation}
\mathcal{W} = \left\{ M_k :\frac{\max_l(p(M_l|X))}{p(M_k|D)}\leq K \right\},
\end{equation} 
where $K$ is a tuning parameter. Occam's window is the set of all possible models within a reasonable Bayes factor from the best model under consideration. The summation in (\ref{equation::BMA}) is then replaced with a summation over the set $\mathcal{W}$. 
\subsubsection{Averaging the co-clustering matrices}
We can form the Bayesian model-averaged co-clustering matrix (BMAC) by taking the set of co-clustering matrices $S_{\mathcal{W}}$ and averaging, weighting by their ML:
\begin{equation}
S_{BMAC} = \frac{p(X|\mathcal{M}_m)S_{m}}{\sum_{l\in\mathcal{W}}p(X|\mathcal{M}_l)}.
\end{equation}
The BMA of the variable set can be found in the same way by averaging over the weighted variable sets for each model:
\begin{equation}
\mathcal{F}_{BMA} = \frac{p(X|\mathcal{M}_m)\mathcal{F}_{m}}{\sum_{l\in\mathcal{W}}p(X|\mathcal{M}_l)},
\end{equation}
where we denote by $\mathcal{F}_{m}$ the variable set associated with model $\mathcal{M}_m$.

%%%%%%%%RESULTS1:
\section{Comparisons with the state-of-the-art}\label{sec:verify}
We compare SUGSVarSel to a number of alternative algorithms, and demonstrate the performance of our method in two situations. The first is the $p > n$ paradigm, where the number of variables exceeds the number of observations. The second situation considers $n > p$ for $n = 1000$, while simultaneously considering different proportions of variables being relevant.  In both cases, we consider a variety of scenarios, for which different proportions of the variables are relevant. 
\subsection{Alternative methods for clustering and variable selection}
We compare our method relative to the current state-of-the-art, including methods that do and do not peform variable selection. These include: mclust, a finite mixture model based clustering method \citep{Fraley::2002, mclust::2012, mclust::2016}; clustvarsel, a finite mixture model method with variable selection \citep{Raftery::2006, Maugis::2009, Scrucca::2014}; the original sequential updating and greedy search algorithm \citep{Wang::2011} as implemented in our sugsvarsel R package; and VarSelLCM, a model-based clustering and variable selection approach using the integrated complete-data likelihood \citep{Marbac::2017}.

\subsection{High-dimensional example}
In the first example, we simulate a mixture of $3$ Gaussians with mixture proportions $0.5, 0.3, 0.2$ centred at $(0,0,..,0), (2,2,...,2), (-2,-2,...,-2)$ respectively, each with variance-covariance matrix equal to the identity. The irrelevant variables are simulated from a standard Gaussian. First, we simulate $100$ observations from this model with $200$ variables and explore varying the number of relevant variables.  
\\
\\
When running SUGS and SUGSVarSel we use the same prior specification for both methods and $30$ random orderings of the data. Throughout this article, we always perform $2$ iterations of variable selection in the SUGSVarSel algorithm. To initialise variable selection in SUGSVarSel, we subsample $10\%$ of the variables $20$ times to produce an initial variable selection set. For SUGS we choose the partition with maximal PML (as advised in the original SUGS paper by \citealt{Wang::2011}), while for SUGSVarSel we select the result with maximal ML. Prior choices for SUGS and SUGSVarSel can be found in the appendix. For mclust and clustvarsel, we find the appropriate number of clusters using a sequential search up to a maximum of 9 possible clusters. We then use then Bayesian Information Criterion (BIC) to select an appropriate model \citep{Schwarz::1978}. For VarSelLCM we run the algorithm up to a maximum of 9 possible clusters and select an appropriate model using the Maximum Integrated Complete-data Likelihood (MICL) \citep{Marbac::2017, Matthieu:2018}. All methods are run in serial for fair comparison.
\\
\\
Results are presented in Tables \ref{tab1} -- \ref{tab4}.  In all tables, we provide runtimes for each of the methods, indicate the proportion of relevant and irrelevant variables that each method correctly identified (for methods without variable selection this is reported as 1 for relevant and 0 for irrelevant variables), and report the adjusted Rand index \citep{rand::1971, Hubert::1985} between the clustering produced and the truth. We repeat all methods for $10$ different random realisation of the datasets to produce a distribution of scores. We report the median scores, along with the upper and lower quartiles. 

\begin{table}[H]
	\caption{High-dimensional simulation example where $100$ observations are simulated from a Gaussian mixture distribution with $3$ components and 200 variables, in which 50\% of variables are relevant.}\label{tab1}
	\centering 
	\begin{tabular}{c c c c c} % centered columns (4 columns)
		\hline %inserts double horizontal lines
		Method & Time (secs) & Correct  & Correct  & ARI   \\ % inserts table
		 & &  relevant  &  irrelevant  &    \\  % inserts table
		 & & variables &   variables &    \\ [0.5ex] % inserts table
		%heading
		\hline\hline % inserts single horizontal line
		mclust & $<1$ & 1 &  0 & 1 [1, 1] \\
		\hline %inserts single line
		clustvarsel & 6780 & 0.45 & 1 & 1\\
		\hline% inserts single horizontal line
		SUGS & 0.92 [0.90, 0.97] & 1 &  0 & 0.955[0.90, 0.97] \\
		\hline %inserts single line
		SUGSVarSel & 24.6 [23.8, 24.9] & 1 [1, 1] & 1 [1, 1] & 1 [1, 1]\\
		\hline %inserts single line
		VarSelLCM & 580.7 [574.8, 587.9]  & 1 [1, 1] &  1 [1, 1] & 1 [1, 1] \\
		\hline %inserts single line
	\end{tabular}
\end{table}
\begin{table}[H]
	\caption{High-dimensional simulation example where $100$ observations are simulated from a Gaussian mixture distribution with $3$ components and 200 variables, in which 25\% of variables are relevant.}\label{tab2}
	\centering 
	\begin{tabular}{c c c c c} % centered columns (4 columns)
		\hline %inserts double horizontal lines
		Method & Time (secs) & Correct  & Correct  & ARI   \\ % inserts table
		 & &  relevant  &  irrelevant  &    \\  % inserts table
		 & & variables &   variables &    \\ [0.5ex] % inserts table
		%heading
		\hline\hline % inserts single horizontal line
		mclust & $<1$ & 1 &  0 & 1 [1, 1] \\
		\hline %inserts single line
		clustvarsel & 3825.52  &0.02 &  0.673 & 0 \\
		\hline% inserts single horizontal line
		SUGS & 2.07 [1.89, 2.16] & 1 &  0 & 0.78 [0.72, 0.84] \\
		\hline %inserts single line
		SUGSVarSel & 21.9 [21.9, 22.1] & 1 [1, 1] & 1 [1, 1] & 1 [1, 1] \\
		\hline %inserts single line
		VarSelLCM & 508.0 [496.3, 517.0] & 1 [1, 1] & 1 [1, 1] & 1 [1, 1] \\
		\hline %inserts single line
	\end{tabular}
\end{table}
\begin{table}[H]
	\caption{High-dimensional simulation example where $100$ observations are simulated from a Gaussian mixture distribution with $3$ components and 200 variables, in which 10\% of variables are relevant.}\label{tab3}
	\centering 
	\begin{tabular}{c c c c c} % centered columns (4 columns)
		\hline %inserts double horizontal lines
		Method & Time (secs) & Correct  & Correct  & ARI   \\ % inserts table
		 & &  relevant  &  irrelevant  &    \\  % inserts table
		 & & variables &   variables &    \\ [0.5ex] % inserts table
		%heading
		\hline\hline % inserts single horizontal line
		mclust & $<1$ & 1 &  0 & 0 [0, 0] \\
		\hline %inserts single line
		clustvarsel & 6459.88 & 0.1 &  0.772 & 0 \\
		\hline% inserts single horizontal line
		SUGS & 5.02 [4.76, 5.23] & 1 &  0 & 0.18 [0.13, 0.21] \\
		\hline %inserts single line
		SUGSVarSel & 19.7 [19.5, 19.9] & 1 [1, 1] &  1 [1, 1] & 1 [1, 1] \\
		\hline %inserts single line
		VarSelLCM & 523.3 [521.3, 527.7]  & 1 [1, 1] & 1 [1, 1] & 1 [1, 1] \\
		\hline %inserts single line
	\end{tabular}
\end{table}
\begin{table}[H]
	\caption{High-dimensional simulation example where $100$ observations are simulated from a Gaussian mixture distribution with $3$ components and 200 variables, in which 5\% of variables are relevant.}\label{tab4}
	\centering 
	\begin{tabular}{c c c c c} % centered columns (4 columns)
		\hline %inserts double horizontal lines
		Method & Time (secs) & Correct  & Correct  & ARI   \\ % inserts table
		 & &  relevant  &  irrelevant  &    \\  % inserts table
		 & & variables &   variables &    \\ [0.5ex] % inserts table
		%heading
		\hline\hline % inserts single horizontal line
		mclust & $<1$ & 1 & 0 & 0 [0, 0] \\
		\hline %inserts single line
		clustvarsel  & 178.66 & 0.1 &  0.01 & 0 \\
		\hline% inserts single horizontal line
		SUGS & 6.30 [6.07, 10.11] & 1 &  0 & 0.04 [0.02, 0.05] \\
		\hline %inserts single line
		SUGSVarSel & 19.9 [19.7, 20.5] & 1 [1, 1] & 1 [1, 1] & 1 [1, 1] \\
		\hline %inserts single line
		VarSelLCM & 996.8 [959.7, 1084.3] & 1 [1, 1] &  1 [1, 1]   & 1 [1, 1] \\
		\hline %inserts single line
	\end{tabular}
\end{table}
It is evident that methods that do not perform variable selection such as mclust and SUGS perform poorly when there are many irrelevant variables. The performance of clustvarsel here seems volatile and performs poorly at correctly selecting relevant features. VarSelLCM and SUGSVarSel are competitive in terms variable selection and clustering. However, VarSelLCM requires an exhaustive search over the number of clusters, which makes this method computationally costly to apply when the number of clusters is not known. SUGSVarSel outperforms all variable selection and clustering methods in terms of speed, while also automatically inferring the number of clusters in the data. We proceed to evaluate the performance of SUGSVarSel on large simulated datasets.
\subsubsection{Increasing the number of observations}
We simulate the same distribution as before, but instead sample $1000$ observations and only $100$ variables and the irrelevant variable are simulated from a standard Gaussian distribution. All priors are the same as in the previous analysis and we sub-sample $10\%$ of the variables $10$ times to produce an initial variable selection set. We repeat SUGS and SUGSVarSel for $10$ random orderings of the data. We compare the scalable methods mclust, SUGS, SUGSVarSel and VarSelLCM, where $25\%, 10\%, 5\%$ of the variable are relevant. For SUGS we choose the partition with maximal PML, while for SUGSVarSel we select the result with maximal ML. For VarSelLCM we run the algorithm for possible number of clusters 1 through 4 and select an appropriate model using the MICL, as previously.  Results are presented in Tables \ref{tab5}--\ref{tab7}. 
\begin{table}[H]
	\caption{Simulation example where $1000$ observations are simulated from a Gaussian mixture distribution with $3$ components and 100 variables, in which 25\% of variables are relevant.}\label{tab5}
	\centering 
	\begin{tabular}{c c c c c} % centered columns (4 columns)
		\hline %inserts double horizontal lines
		Method & Time (secs) & Correct  & Correct  & ARI   \\ % inserts table
		 & &  relevant  &  irrelevant  &    \\  % inserts table
		 & & variables &   variables &    \\ [0.5ex] % inserts table
		%heading
		\hline\hline % inserts single horizontal line
		mclust & 11.2 [10.9, 11.6] & 1 & 0 & 0 [0, 0]\\
		\hline% inserts single horizontal line
		SUGS & 3.4 [3.1, 3.6] & 1 & 0  &  0.98 [0.97, 0.98]\\
		\hline %inserts single line
		SUGSVarSel & 31.2 [30.7, 31.8] & 1 [1, 1] & 1 [1, 1] & 1 [1, 1]\\
		\hline %inserts single line
		VarSelLCM  & 3596.8 [2639.5, 7537.7]  & 1 [1, 1]   & 1 [1, 1]   & 1 [1, 1]   \\
		\hline %inserts single line
	\end{tabular}
\end{table}
\begin{table}[H]
	\caption{Simulation example where $1000$ observations are simulated from a Gaussian mixture distribution with $3$ components and 100 variables, in which 10\% of variables are relevant.}\label{tab6}
	\centering 
	\begin{tabular}{c c c c c} % centered columns (4 columns)
		\hline %inserts double horizontal lines
		Method & Time (secs) & Correct  & Correct  & ARI   \\ % inserts table
		 & &  relevant  &  irrelevant  &    \\  % inserts table
		 & & variables &   variables &    \\ [0.5ex] % inserts table
		%heading
		\hline\hline % inserts single horizontal line
		mclust & 11.0 [10.7, 11.4] & 1 & 0 & 0 [0, 0] \\
		\hline% inserts single horizontal line
		SUGS & 5.1 [4.9, 5.3] & 1 & 0 & 0.01 [0.01, 0.04] \\
		\hline %inserts single line
		SUGSVarSel & 33.3 [33.0, 33.8]   & 1 [1, 1] & 1 [1, 1] & 0.90 [0.80, 0.97] \\
		\hline %inserts single line
		VarSelLCM & 1938.5 [1852.3, 1973.9] & 1 [1, 1] & 1 [1, 1]  & 0.997 [0.994 0.997] \\
		\hline %inserts single line
	\end{tabular}
\end{table}
\begin{table}[H]
	\caption{Simulation example where $1000$ observations are simulated from a Gaussian mixture distribution with $3$ components and 100 variables, in which 5\% of variables are relevant.}\label{tab7}
	\centering 
	\begin{tabular}{c c c c c} % centered columns (4 columns)
		\hline %inserts double horizontal lines
		Method & Time (secs) & Correct  & Correct  & ARI   \\ % inserts table
		 & &  relevant  &  irrelevant  &    \\  % inserts table
		 & & variables &   variables &    \\ [0.5ex] % inserts table
		%heading
		\hline\hline % inserts single horizontal line
		mclust  & 11.4 [11.2, 15.7] & 1 & 0 & 0 [0, 0] \\
		\hline% inserts single horizontal line
		SUGS  & 6.3 [5.6, 11.1] & 1 & 0 & 0 [0, 0]\\
		\hline %inserts single line
		SUGSVarSel  & 60.8 [59.8, 64.2] & 1 [1, 1] & 1 [0.99, 1] & 0.78 [0.54, 0.92] \\
		\hline %inserts single line
		VarSelLCM & 2688.8 [2588.9, 2878.6] & 1 [1, 1] & 1 [1, 1] & 0.943 [0.931, 0.945]\\
		\hline %inserts single line
	\end{tabular}
\end{table}
SUGSVarSel and VarSelLCM produce high quality answers in all situations but SUGSVarSel is 2 orders of magnitude faster. However, to alleviate the computational burden we searched up to a maximum of 4 clusters in VarSelLCM, providing it with an easier opportunity to produce high quality clusterings. In applications to real data this would have to be much larger, adding considerably to computational time, whereas the inference of the number of clusters is automatic in SUGSVarSel. 
\subsection{Advantages of Bayesian model averaging}
\begin{figure}[H]
	\includegraphics[width=0.5\linewidth]{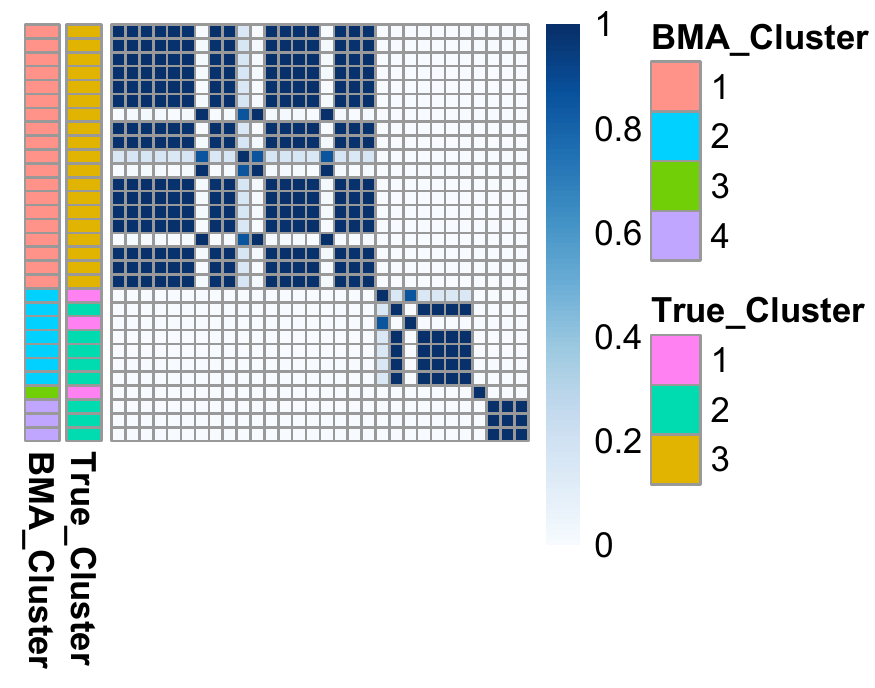}
	\centering
	\caption{A heatmap of the BMA co-clustering matrix, where dark blue indicates the probability of being in the same cluster is 1 and white indicates a probability of 0 of belonging to the same cluster. The component annotation bar indicates the true component labels and the cluster annotation bar indicates the clustering obtained from summarising the BMA co-clustering matrix.}\label{bmaFig}
\end{figure}

%Bayesian model averaging allows us to capture the uncertainty in our models and this is particularly useful when our independence assumptions are not met. When our modelling assumptions are met we find that a single particular model is most likely and dominates the averaging. However, when no particular model has a dominate marginal likelihood we find averaging produces more robust and allows to explore our model uncertainty. 
%\\
%\\
As an example, we simulate a dataset with $30$ observations from a mixture of 3 Gaussians, where two of the Gaussians are isotropic and centred (2,2) and (-3,-3) respectively, each with mixing weights $0.4$. The third component has mixture weight $0.2$ and is centered at (-3,4) but the covariance matrix is 2 on the diagonals and 1 on the off diagonals, violating our independence assumption. We additionally include 2 components of irrelevant variables generated from standard Gaussians. Our prior specifications are set as in the previous section. Simply using the ML to pick a partition results in an ARI of 0.635 between the clustering produced and the truth. However, we can also perform BMA and then summarise our co-clustering. We applied hierarchical clustering with average linkage to compute a clustering, which has previously be applied to posterior similarity matrices \citep{Medvedovic::2004, fritsch::2009, Liverani::2015} (see appendix for complete details). This clustering then produces an ARI of 0.875. The heatmap of the co-clustering matrix is Figure~\ref{bmaFig}, providing a visualisation of the uncertainty in the clustering.

\newpage
%%%%%%%%RESULTS2:

\section{Applications to cancer subtyping}
\subsection{Application to Leukaemia Dataset}
In this section, we apply SUGSVarSel to real biological datasets. The first is a well-studied genomic clustering problem: the separation of acute myeloid leukaemia (AML) and the B/T-cell subtypes of acute lymphoblastic leukemia (ALL) samples on the basis of microarray transcriptomic data.  We use the dataset described by \citet{Golub::1999}, which comprises 38 samples, 27 of which are ALL ($8$ T-cell and $19$ B-cell related), and 11 of which are AML cases. Initial preprocessing is performed as in \cite{Dudoit::2002}, which reduces the dimension of the dataset from 6,817 to 3,051 genes.  In \cite{Dudoit::2002}, a further dimension reduction step is performed that makes use of the AML and ALL class labels, so that only those genes that have a high ratio of their between-class to within-class sums of squares are retained.  Here we instead wish to adopt a completely unsupervised approach, so that we may use the known ALL-AML class label in order to validate our results.  

We select the 200 most variable genes and then normalise, so the expression values for each gene are mean-centred at 0 with variance 1. $200$ genes were chosen because this led to good predictive performance in previous analysis of these data \citep{Golub::1999,Dudoit::2002}. We then apply SUGSVarSel to the resultant dataset.  We sub-sample $10\%$ of the variables $20$ times to produce an initial variable selection set, and run the algorithm for $100$ random orderings. We adopt our default priors and summarise the output using BMA.  A final summary clustering is obtained by performing hierarchical clustering with average linkage \citep{fritsch::2009}. We use the ARI to compare our results to the truth (of 3 classes) and repeat the process 10 times and report the average results.  

\begin{figure}[H]
	\centering
%	\begin{subfigure}[t]{0.5\textwidth}
%		\centering
%		\includegraphics[width=6.5cm]{corplotGolub.pdf}
%		\caption{}
%		\label{figure::corplotGolub}
%	\end{subfigure}%
%	\begin{subfigure}[t]{0.5\textwidth}
%		\centering
%		\includegraphics[width=6.5cm]{allcorGolub.pdf}
%		\caption{}
%		\label{figure::allcorGolub}
%	\end{subfigure}
%	\vspace{1cm}
%	\begin{subfigure}[t]{0.5\textwidth}
%		\centering
		\includegraphics[width=0.6\linewidth]{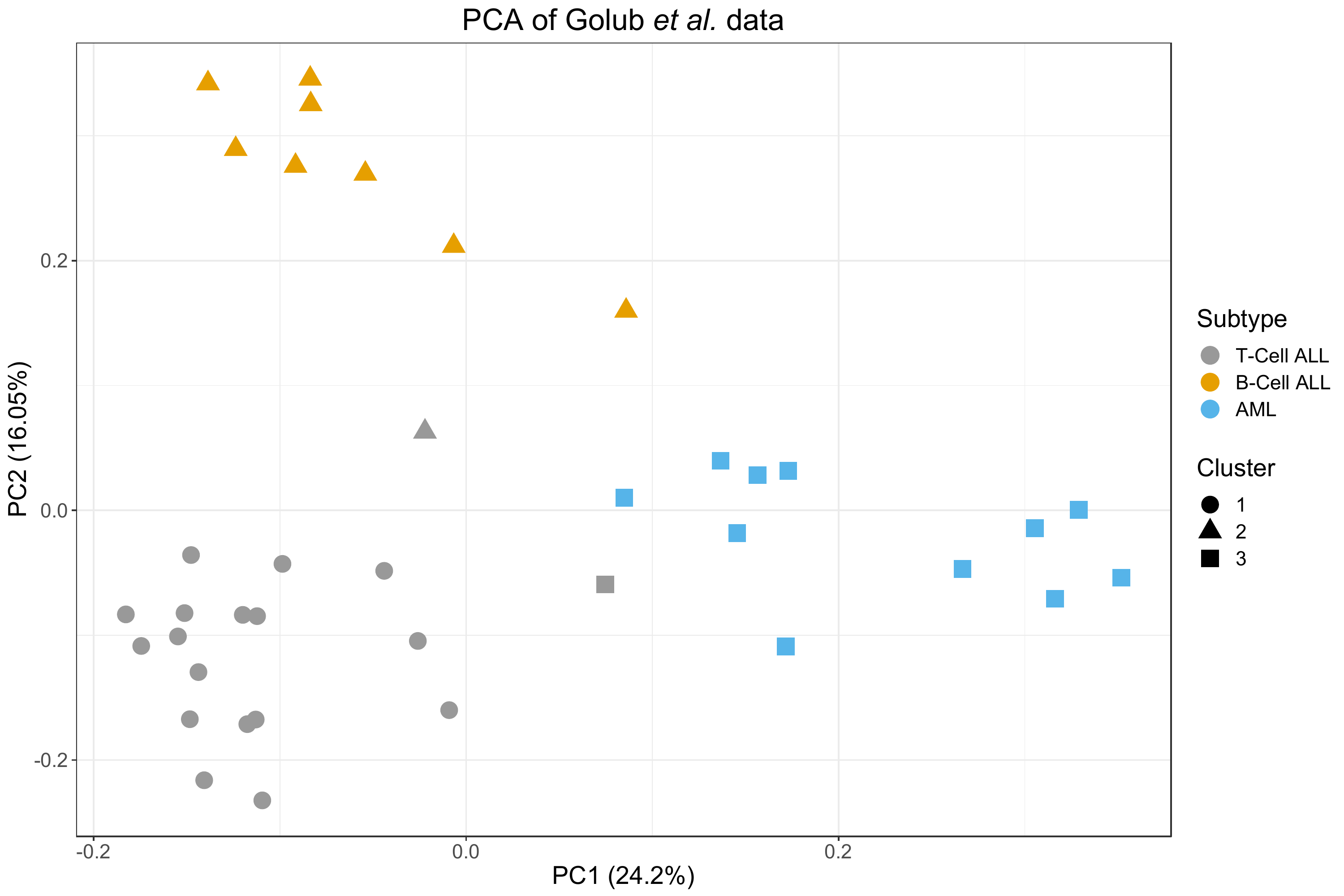}
%		\caption{}
%		\label{figure::pcaGolub} 
%	\end{subfigure}
	\caption{A PCA plot of the microarray expression data of 38 patients from the \cite{Golub::1999} dataset, using the 200 most variable genes. The different symbols indicate the clustering produced by the SUGSVarSel algorithm after summarising the BMA co-clustering matrix using hierarchical clustering with average linkage. The colours indicate the annotated sub-types.}
	\label{fig::GOLUB}
\end{figure}

Results are illustrated in Figure \ref{fig::GOLUB}.  The final clustering result provides an ARI of $0.831$, which is in line with previous analyses preformed on this dataset \citep{Golub::1999, Dudoit::2002}. The algorithm selects a total of 92 genes, including TCL1, TCRB, IL8, EPB72, IL7R, TCRG, NFIL6, which are all known to be associated with leukaemia \citep{Pekarsky::2001, Van::2004,Kuett::2015,Chen::2010, Shochat::2011, Natsuka::1992}. A full list of the selected genes (including their descriptions) can be found in the appendix.  %, for which we plot the correlation matrix between patients showing that we achieve a clear separation between the AML-ALL patients, and additionally the B-cell and T-cell subgroups of ALL. 
The advantage of our analysis over other methods is that we did not need to specify the number clusters - the algorithm automatically inferred 3 clusters in the data, which have excellent correspondence to the known classes of AML and ALL, as well as the 2 ALL subgroups.

\begin{figure}[H]
	\centering
	\includegraphics[trim=4cm 4cm 4cm 4cm,clip=true, width=0.7\linewidth]{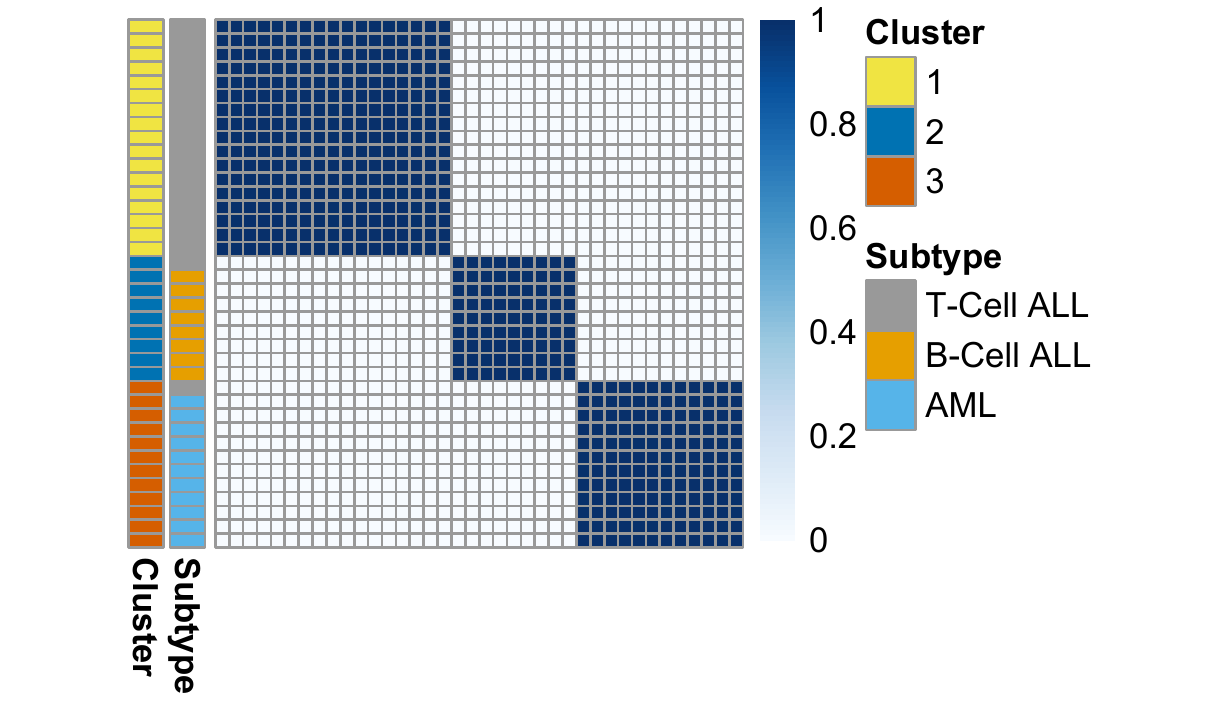}
	\caption{A heatmap of the BMA co-clustering matrix for the 38 patients, when applying SUGSVarSel, demonstrating the added benefit of visualising uncertainty. The annotation bars of the left indicate the correspondence between the clusters and the subtypes.}
	\label{fig::BMAGolub}
\end{figure}

To assess the importance of variable selection, we also apply mclust and the original SUGS algorithm to the data. We run the mclust algorithm performing a systematic search to select the number of clusters, up to a maximum of 9, and select the number of cluster which maximises the BIC. This criterion selects 3 clusters and clustering produced gives an adjusted Rand index of $0.627$ - the inclusion of irrelevant variables has led to reduced cluster quality. We run SUGS using our default prior choices and using the PML criterion to select a clustering. The algorithm was run for $100$ random ordering and we repeated the process $10$ times, reporting an average ARI of $0$. The lack of variable selection renders SUGS unable to produce a meaningful clustering. In Figure \ref{fig::BMAGolub}, we visualise the BMA co-clustering matrix for these data when applying the SUGSVarSel algorithm.
%%%%%%%%RESULTS2B:
\subsection{Application to TCGA breast cancer dataset}
We demonstrate SUGSVarSel on a further genomics dataset. We analyse an expression dataset for breast cancer tumour data from The Cancer Genome Atlas (TCGA)\citep{Cancer::2012}, which we pre-process in the same way as in \cite{Lock::2013}. The processed expression dataset comprises $348$ tumours with $645$ genes, of which $14$ belong to the PAM50 (Prediction Analysis of Microarray) group of genes \citep{Parker::2009}. %The PAM50 group of genes are relevant because they form part of a standard tumour profiling test, which can inform whether chemotherapy or hormonal therapy might be beneficial for the patient.

\begin{figure}[H]
	\includegraphics[width=\linewidth]{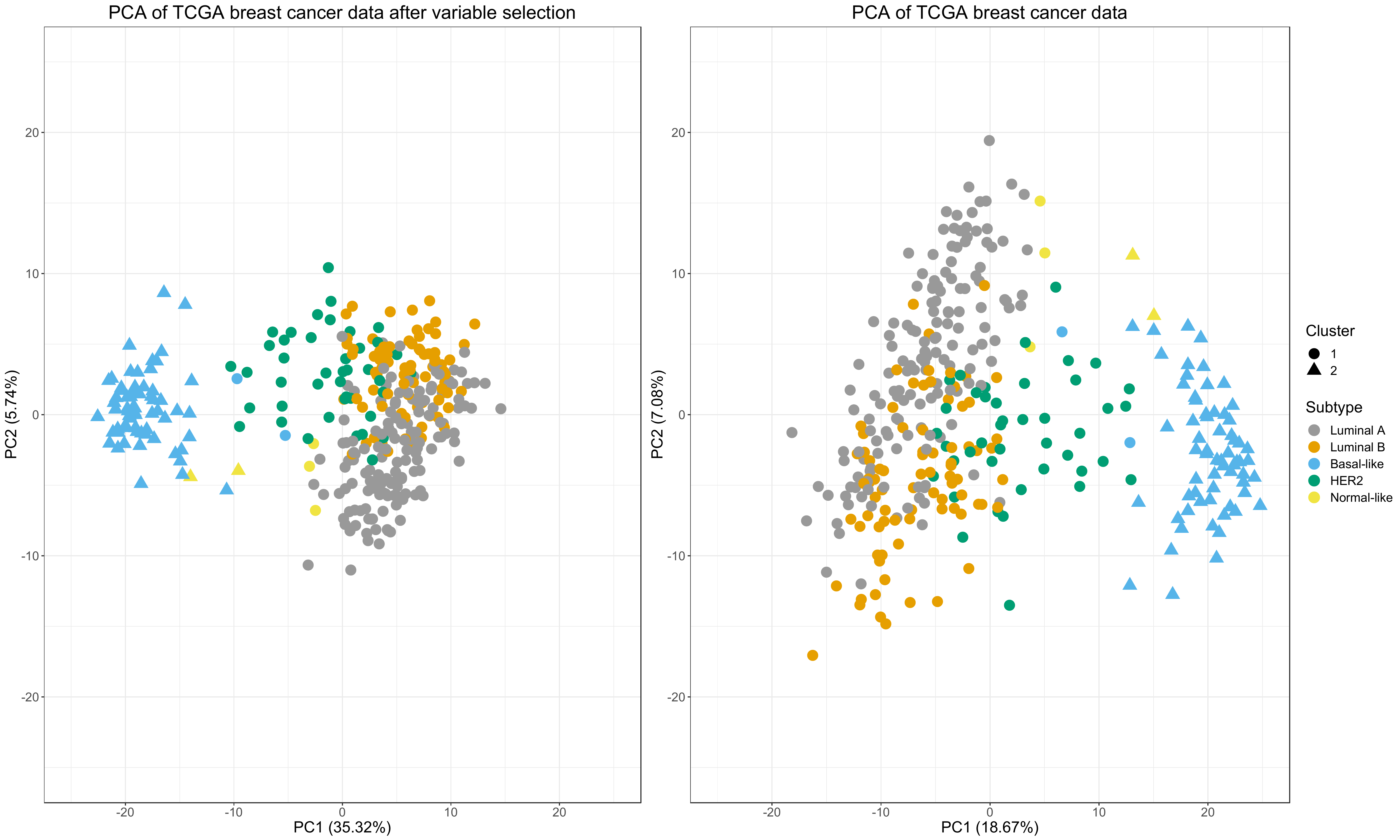}
	\centering
	\caption{PCA plot of the TCGA breast cancer data, where clusters identified by SUGSVarSel are indicated by shape and subtypes by colour. The left PCA plot demonstrates smaller and tighter clusters using only the variables that remained after variable selection. In the right hand plot all variable were used to produce the plot.}
	\label{figure::pcabreastcancer}
\end{figure}

Analysis was performed in the following way. We first standardise our data so that each column is mean-centred with variance 1. We then subsample $10\%$ of the variables $64$ times to produce an initial variable set. We then apply the SUGSVarSel algorithm with default settings. We summarise our output by performing BMA and then hierarchical clustering with average linkage.
\\
\\
SUGSVarSel reveals two clusters in the dataset, the second of which is significantly associated with Basal-like tumours (Fisher test, $p < 0.0001$). The algorithm selects $245$ variables to discriminate between the groups. We perform PCA before and after variable selection to demonstrate that the reduced variable set produces more separable and therefore more interpretable clusters (Figure~\ref{figure::pcabreastcancer}).
Furthermore, the algorithm selected $13$ out of a total of $14$ of the PAM50 genes, which is significantly better than random (Fisher Test, $p < 0.0001$). 
\\
\\
There is perhaps concern that variable selection could remove relevant genes for clustering, in the situation where we have a highly informative set of variables.
We consider the following task to cluster the breast cancer genes using the PAM50 genes from the total unprocessed dataset (that is without the filtering of \cite{Lock::2013}), of which there are $48$. We apply the SUGSVarSel algorithm in identical fashion to before, sub-sampling $10\%$ of the variables $4$ time to produce an initial variable set. We obtain $5$ clusters which correspond well to the different breast cancer subgroups.
\begin{figure}[H]
	\includegraphics[width=\linewidth]{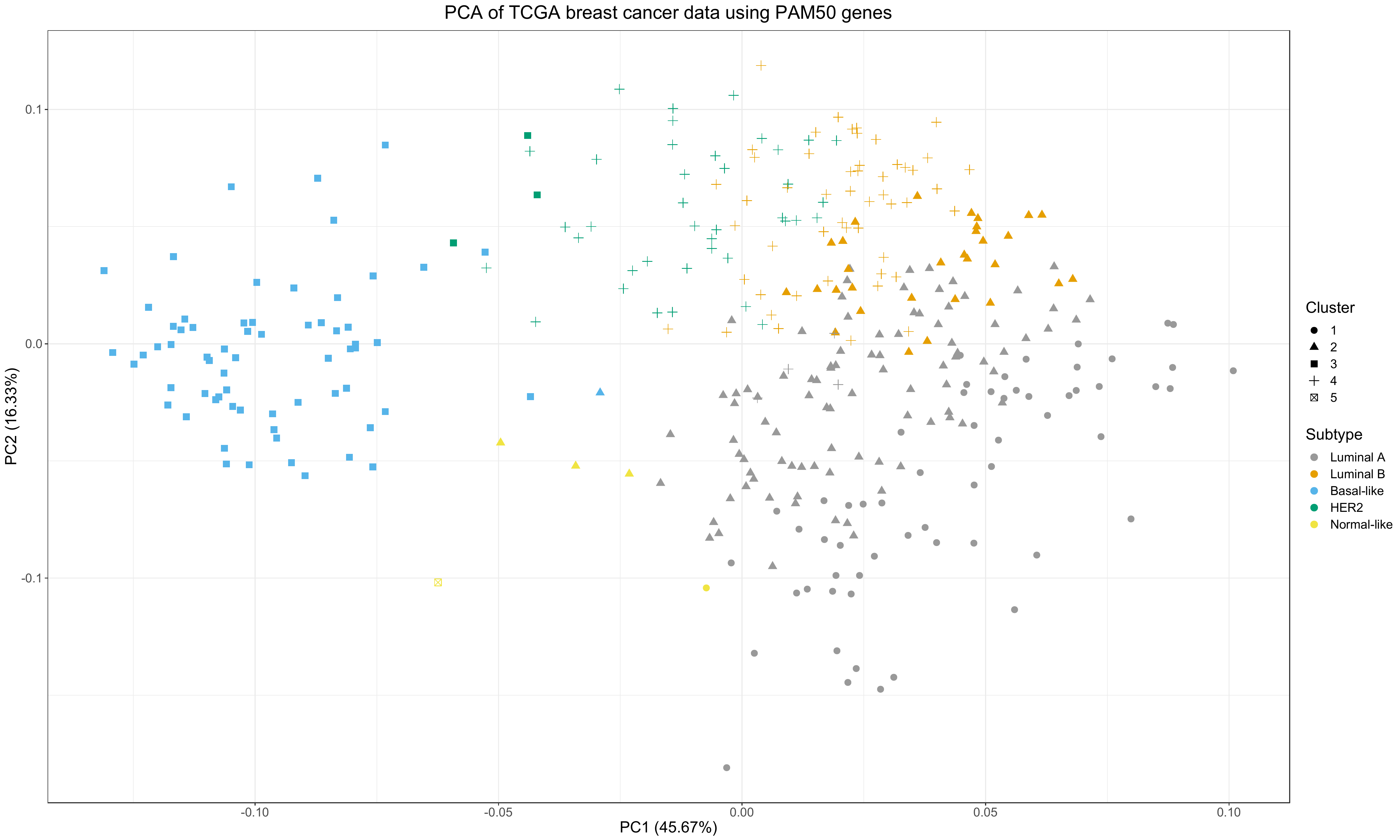}
	\centering
	\caption{PCA plot of the TCGA breast cancer data using 48 of the PAM50 genes, where clusters produced by SUGSVarSel are indicated by shape and subtypes by colour. }
	\label{figure::pcapam50}
\end{figure}
Cluster A is associated with Luminal A cancers, cluster B is associated with Luminal cancers, cluster C with basal-like tumours, cluster D contains mostly HER2 type breast cancers (chi-squared $p<0.0001$). Thus, hardly surprisingly, the cluster produce on the PAM50 data coincide well with the PAM50 subgroups. Furthermore, $87.5\%$ of the genes were selected which is more than we expect given our prior, telling us this was a highly informative set of genes.
\\
\\
The clusterings shown in Figures \ref{figure::pcabreastcancer} and \ref{figure::pcapam50} demonstrate that the variables we use for clustering are critically important. The two different pre-filtering choices led to results of varying quality and biological meaning. This is strong evidence in support of model-based variable selection rather than ad-hoc preprocessing.	
%%%%%%%%RESULTS3:

\section{Pan-cancer proteomic characterisation}
\begin{figure}[H]
	\includegraphics[width=10cm]{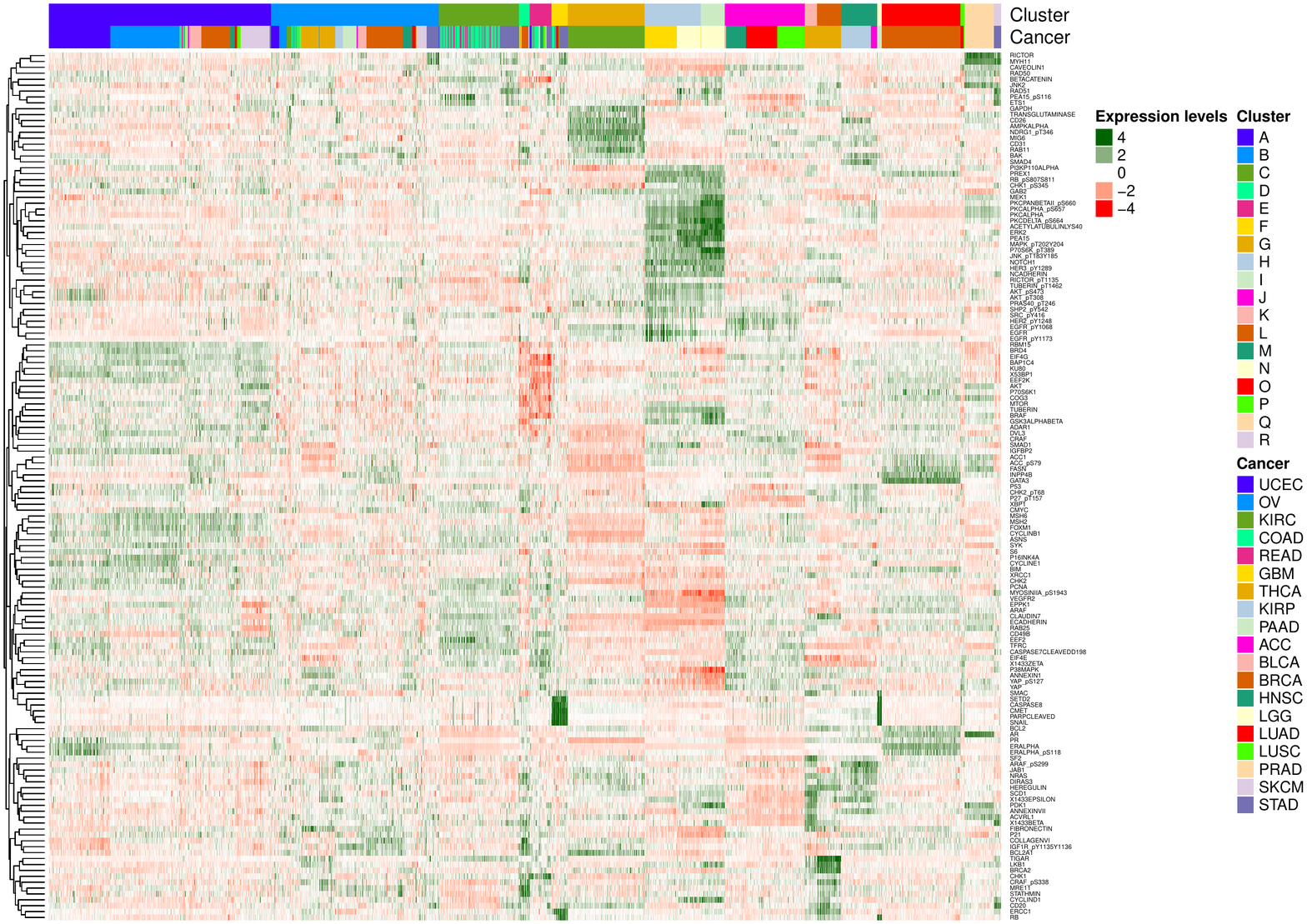}
	\centering
	\caption{A heatmap of the expression data using the clustering produced by the SUGSVarSel algorithm applied to the pan-cancer TCPA dataset. The annotation bars on the top of plot indicate the different cancers and clusters.}\label{figure::TCPAsugsheat}
\end{figure}

\begin{figure}[H]
 	\includegraphics[width=10cm]{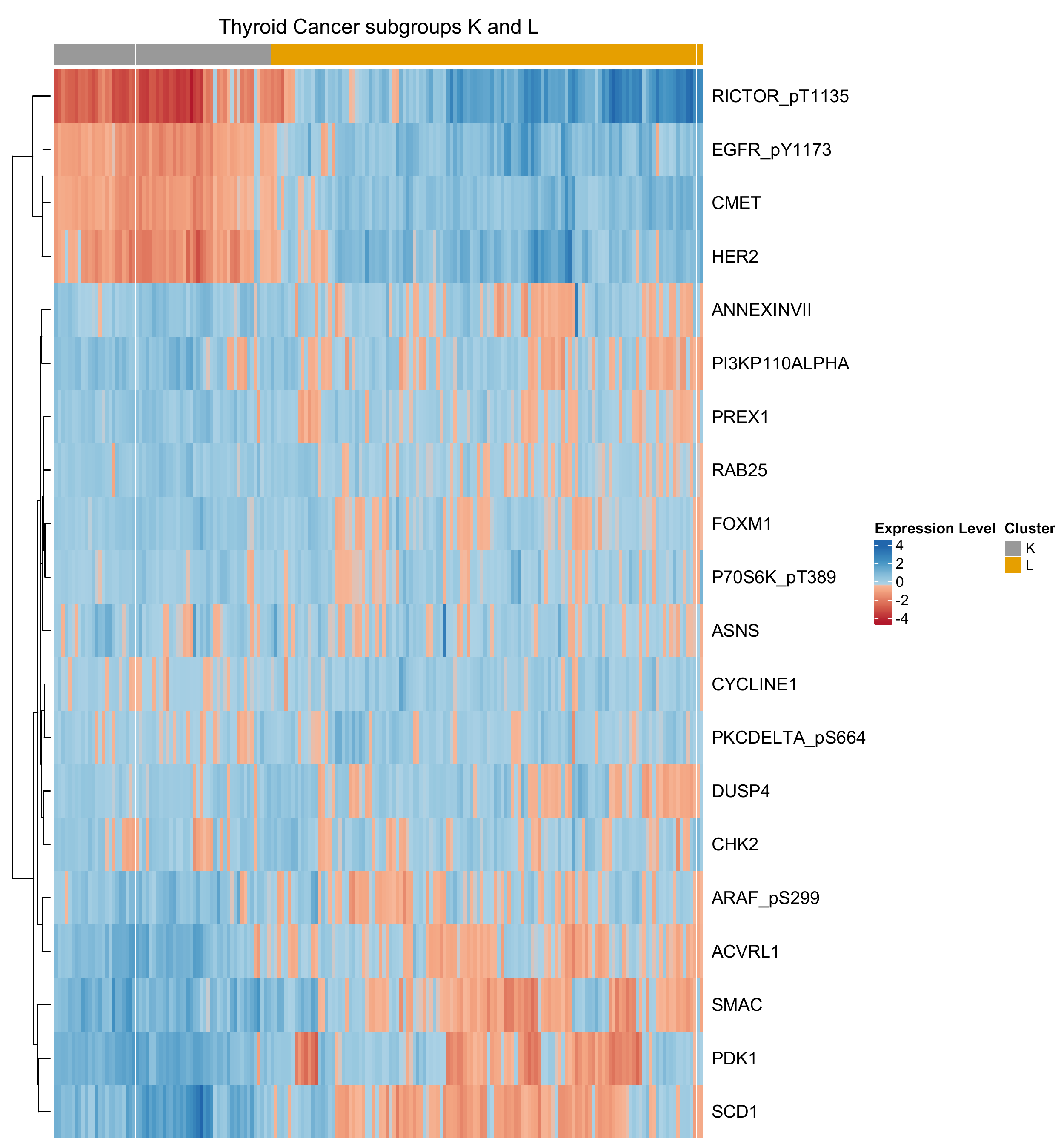}
 	\centering
 	\caption{A heatmap of the expression TCPA data for the thyroid subgroups. We have plotted the expression for only the top $20$ proteins which are significantly different between clusters K and L.}
 	\label{figure::thyroidHeat}
\end{figure}

In this section we apply our method to The Cancer Proteome Atlas (TCPA) datasets \citep{Li::2013, Akbani::2014, Stadler::2017}. The dataset contains a large number of tumours and cell line samples with protein expression levels generated using reverse-phase protein arrays (RPPAs). Our method allows us to perform a number of tasks on this data; in particular, for each cancer we can detect possible subgroups and the relevant proteins which discriminate these subgroups. We can also perform a pan-cancer analysis to explore the differences and similarities between cancers. Pan-cancer studies can unravel inter-cancer relationships which are important for developing new clinical targets \citep{Weinstein::2013, Uhlen::2017, Berger::2018, Hoadley::2018}. Recent pan-cancer analyses have suggested that cancers should be classified based on their molecular signatures rather than tissue of origin \citep{Berger::2018, Hoadley::2018} and this motivates our analysis.
\\
\\
As is usual with this data there are irrelevant variables so methods that do not perform variable selection such as mclust and SUGS are ill-suited. Furthermore, there is little \textit{a priori} knowledge about the number of clusters and so methods such as VarSelLCM and clustvarsel which require an exhaustive search of the number of clusters are inappropriate. To perform the analysis on all cancer sets would be prohibitively slow for the slowest of analysis methods.
\\
\\
The TCPA datasets contain data on 19 cancer types and the description of these cancers can be found in the appendix. The total dataset consists of over 5000 tumour samples with only a few samples for some cancers and hundreds of samples for others and several hundreds of proteins. The merged PAN-Can 19 level 4 dataset is used in the following analysis, since it is appropriate for multiple disease analysis. More information about the data can be found here \url{http://tcpaportal.org/tcpa/}, where the data itself can also be downloaded. In addition, we standardise the expression levels for each protein so that they are zero-centred with unit variance.
\\
\\
Table~\ref{numPerCancer} demonstrates the number of cases for each cancer type:
\begin{table}[ht]
	\caption{A table indicating the different cancer types and the number of observations from each of those cancers.}\label{numPerCancer}
\setlength{\tabcolsep}{4pt} % Default value: 6pt
\renewcommand{\arraystretch}{0.8} % Default value: 1
	\centering
	\begin{tabular}{rrrrrrrrrrrr}
		\hline
		ACC & BLCA & BRCA & COAD & GBM & HNSC & KIRC & KIRP & LGG & LUAD \\
				\hline
		46 & 127 & 820 & 327 & 205 & 203 & 445 & 208 & 257 & 234 \\
				\hline
		LUSC & OV & PAAD & PRAD & READ & SKCM & STAD & THCA & UCEC & &\\  
		\hline
		192 &411 & 105 & 164 & 129 & 207 & 299 & 374 & 404 & &\\
	\end{tabular}
\end{table}
\\
We only keep proteins which have been measured on all cancers, which total $217$ and so our dataset has a total of $5157$ tumour samples with $217$ variables.
We apply SUGSVarSel to this data by first sub-sampling $10\%$ of the variables $43$ (a fifth of the total number of variables) times. Using the same priors as in previous analysis we analyses this data using the SUGSVarSel algorithm, running the algorithm for $50$ random orderings, thus exploring a total of $2150$ models. We summarise the BMA clustering using hierarchical clustering with average linkage. The summarised clustering contains $60$ clusters, however many of these clusters contain only a few observations. Reassuringly there are $18$ clusters with more than $20$ observations and we focus on these for our analysis. A table summarising the clusters can be found in the appendix. Figure \ref{figure::heatTCPA} summarises the relationship between the cancer types and SUGSVarSel clusterings.
\begin{figure}[H]
	\includegraphics[trim=4cm 4cm 4cm 4cm,clip=true, width=0.75\linewidth]{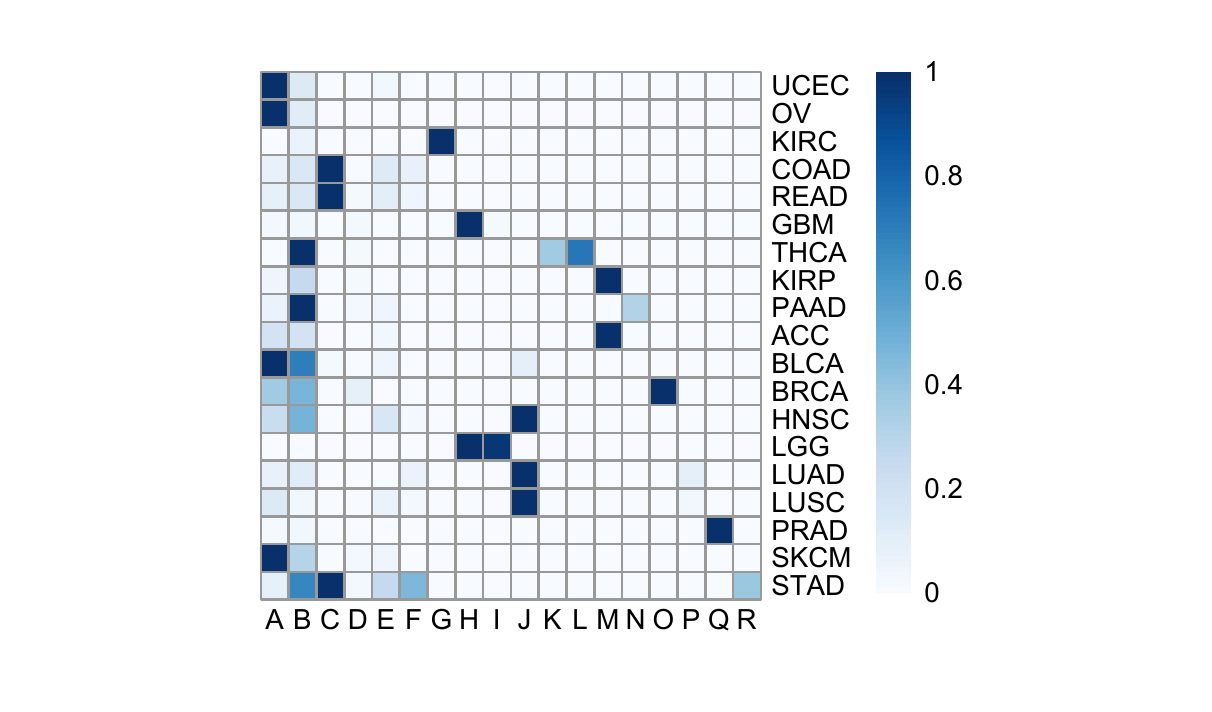}
	\centering
	\caption{A heatmap indicating the correspondence between clusters produced by the SUGSVarSel algorithm and the different cancer types. }
	\label{figure::heatTCPA}
\end{figure}
In addition, in Figure~\ref{figure::TCPAsugsheat} we plot a heatmap of the data with the clustering identifed by SUGSVarSel, using only the proteins selected by the algorithm.
It is rare that a cancer associates with a single cluster, however there are evident relationships between cancers and clusters. Cluster A contains predominately womens' cancers (OV, UCEC, BRCA), while cluster B contains a large spread of cancers. Clusters C, E and F contain the cancers of the digestive tract (STAD, COAD and READ). Cluster D contains a subgroups of breast cancers (BRCA), while cluster G contains solely kidney cancer (KIRC). Clusters H and I contain cancers of the brain (LGG, GBM). Cluster J and P contain aero-digestive cancers (HNSC, LUAD LUSC). Thyroid cancer (THCA) is spread across clusters K, L and B, whilst KIRP is predominately found in cluster M. Pancreatic cancer (PAAD) is split across clusters N and B. Cluster O contains the majority of breast cancer patients.  Prostate cancer (PRAD) is dominantly found in Q, while R forms a small cluster of stomach cancers. This is in line with other analyses performed on these data \citep{Akbani::2014, Hoadley::2014, Csenbabaouglu::2016}. A total of $147$ proteins were selected as relevant for clustering.
\\
\\
We now consider an illustrative example. Figure \ref{figure::heatTCPA} indicates that clusters K and L contain only thyroid cancers. It is of biological interest to see what drives the differences between these clusters, as they could define clinically relevant thyroid cancer subtypes. Considering only the $147$ selected proteins, in Figure~\ref{figure::thyroidHeat} we plot the expression profile for the $20$ proteins, with smallest p-value, which are significantly different between clusters K and L (T-test \citep{Welch::1947}, $ p < 0.00001$, using Benjamini-Hochberg correction \citep{Benjamini::1995}) 
 We do not observe an over representation of known thyroid cancers subtypes within each of these clusters (see Table~\ref{table::thyroid}).
\begin{table}[H]
	\caption{A table showing the distribution of 3 different THCA subtypes across the clusters K and L produce from the SUGSVarSel algorithm. Note that this information was not available for all patients.}\label{table::thyroid}
	\centering
	\begin{tabular}{rrr}
		\hline
		& K & L \\ 
		\hline
		Thyroid Papillary Carcinoma - Classical/usual & 31 & 72 \\ 
		Thyroid Papillary Carcinoma - Follicular ($>$= 99\% follicular patterned) & 17 & 25 \\ 
		Thyroid Papillary Carcinoma - Tall Cell ($>$= 50\% tall cell features) & 2 & 6 \\ 
		\hline
	\end{tabular}

\end{table}
%%%%%%%%CONCLUSION:
\section{Conclusion}
\label{sec:conc}
In this article we presented SUGSVarSel, an extension to the SUGS algorithm of \cite{Wang::2011} to allow variable selection. We demonstrated that when irrelevant variables are present the quality of the clustering can be degraded and clusters become more challenging to interpret. SUGSVarSel allows the flexibility of a Bayesian nonparametric approach but inference is considerably faster than using MCMC. Indeed, the SUGSVarSel algorithm infers the number of clusters automatically and performs inference for the Dirichlet process hyperparameter. This is in contrast to most clustering with variable selection methods which require a systematic search over the number of clusters.
\\
\\
Whilst our method is approximate it performs competitively with other commonly used approaches. Furthermore, we take advantage of exploring many models by performing Bayesian model averaging, which is important for exploring uncertainty in our clustering. We remark that model uncertainty and the application of BMA is rarely explored in clustering tasks. We have provided an R package to facilitate dissemination of our method utilising C++ to accelerate intensive computations and parallel processing features to make further computational gains
\\
\\
Application to two cancer transcriptomic datasets show the clear benefit of simultaneously performing variable selection and clustering. We demonstrate that variable selection improves interpretation of these datasets, providing the genes that drive the clustering structure of the data, as well as identifying those that are irrelevant for clustering. We further applied our method to a pan-cancer proteomic dataset for which none of the current model-based clustering and variable selection methods are suitable. SUGSVarSel is able to provide a characterisation of $5,157$ tumour samples, demonstrating clustering relationships across cancer types based on their molecular signature rather than tissue of origin. 
\\
\\
There are a number of ways in which our proposed method could be extended. Firstly, our assumption that variables are conditionally independent given the cluster allocations might be unrealistic for some datasets.  In such cases, more elaborate variable selection methods might be desirable, although this is likely to come at increased computational cost. Furthermore, we have assumed conjugacy throughout, so that the marginal likelihood in Equation \eqref{Lik} may be evaluated analytically.  As noted in the original SUGS paper of \citet{Wang::2011}, one possible way to extend to non-conjugate cases would be to approximate this marginal likelihood, e.g. using a Laplace approximation.
\\
\\
\section*{Acknowledgements}
 O.M.C. is a Wellcome Trust Mathematical Genomics and Medicine student supported financially by the School of Clinical Medicine, University of Cambridge.  L.G. was supported by a BBSRC Strategic Longer and Larger grant (Award BB/L002817/1).  P.D.W.K. was supported by the Medical Research Council grant number MC\_UU\_00002/10.  No conflict of interest declared.

\section{Appendix}
\begin{description}
\item[DP Gaussian mixture with variable selection]
\end{description}
In this appendix, we give specific details for our algorithm in the case of Gaussian mixtures. We note that whenever conjugate priors are chosen all the formulas presented here are available analytically. We specify the mean and covariance matrix associated cluster $k$ by $\theta_{k} = (m_k,\Sigma_k)$,  where $m_k = (m_{k,1},...,m_{k,d})$ and $\Sigma_k$ is diagonal under our independence assumption. The prior on the parameters $p_0(\theta_{k,d})$ is chosen as conjugate normal inverse-chi-squared ($NI\chi^2$) prior, which is a simple reparameterisation of the  normal inverse-gamma prior and is the special case of normal inverse-Wishart prior in one dimension. The specification is as follows:
\begin{equation} \label{equation::prior}
p_0(\theta_{k,d})\sim N\left(m_{k,d}|\mu_0, \frac{\Sigma_{k,d}}{\lambda_0}\right)I\chi^2\left(\Sigma_{k,d}|\nu_0, S_0\right),
\end{equation}
with known hyperparameters $\mu_0,\lambda_0, \nu_0, S_0$.  Updating the prior (\ref{equation::prior}) with the data from observations $1,...,(i-1)$ results in 
\begin{equation}
p(\theta_{k,d}|x_{-i},z_{-i})\sim N\left(m_{k,d}|m^{(i-1)}, \frac{\Sigma_{k,d}}{\lambda^{(i-1)}}\right)I\chi^2\left(\Sigma_{k,d}|\nu^{(i-1)}, S^{(i-1)}\right),
\end{equation}
where the parameter updates are obtained sequentially, through the following equations (dropping the subscript $d$ for clarity) \citep{murphy::2007}:
\begin{equation}
\begin{split}
m^{(i)}_k & = \frac{\lambda_k^{(i-1)}m_k^{(i-1)}+x_i}{\lambda^{(i)}} \\
\lambda^{(i)}_k & = \lambda_k^{(i-1)}+1 \\
\nu_k^{(i)} & =\nu_k^{(i-1)}+1 \\
T_k^{(i)} & = T_k^{(i-1)} +x_i^{2} \\
\nu_k^{(i)}S_k^{(i)} & = T_k^{(i)} - \lambda_k^{(i)}\left(m_{k}^{(i)}\right)^2, 
\end{split}
\end{equation}
where in the case $i=0$ the parameters are given by their specified prior values, except we set $T_k^{(0)} = \nu_0S_0 +\lambda_0\mu_0^2$. The required conditional likelihood is given by a non-central $\mathcal{T}$-distribution. Remembering at the $i^{th}$ iteration we compute with the updated parameters from the previous iteration; that is, at the $(i-1)^{th}$ iteration, the required distribution is 
\[\mathcal{T}\left(\cdot|m^{(i-1)}_k, \nu_k^{(i-1)}, \frac{(1+\lambda_k^{(i-1)})S_k^{(i-1)}}{\lambda_k^{(i-1)}}\right).\]
For a  $NI\chi^2$ prior the marginal likelihood is given by the following equation, (dropping the subscript $d$ for clarity)
\begin{equation}\label{equation::marginalLike}
\int_{\theta_{k}}f(X_{k}|\theta_{k})p_0(\theta_{k})d\theta_{k} = \frac{1}{\pi^{n_k/2}}\frac{\Gamma(\nu_k/2)}{\Gamma(\nu_0/2)}\left(\frac{\lambda_0(\nu_0 S_{0})^{\nu_0}}{\lambda_k(\nu_kS_k)^{\nu_k}}\right)^{1/2}.
\end{equation}
The other required equations have already been given and require simple substitutions.
\\
\begin{description}
\item[Prior Settings for the SUGS and SUGSVarSel high-dimensional example]
\end{description}
Here we state the prior specification for the SUGS and SUGSVarSel algorithms. We let $\mu_0$ be the mean of the observations' data for each variable, $\lambda_0 = 0.01$, $\nu_0$ be the number of variables, $S_0 = 0.2$ for all variables. We let $\hat{\beta} = (0.01, 0.1, 1, 5, 10, 15, 30, 50, 100)^T$ and set the prior to be $\cal{G}$$(1,1)$. In addition, for SUGSVarSel we suppose that \textit{a priori} variables are equally likely to be relevant or irrelevant.
\\
\\
\begin{description}
	\item[Summarising the Bayesian model averaged co-clustering matricies]
\end{description}
\cite{fritsch::2009} propose a method to summarise the posterior similarity matrix of a Bayesian clustering method. We apply their methodology to summarise our Bayesian model averaged co-clustering matrix. They present several method to obtain a clustering by maximising the posterior expected adjusted Rand index. We use the proposed method which obtains clusterings from applying hierarchical clustering with average linkage. An optimal clustering is then obtain by cutting the dendrogram at $0.5$ \citep{fritsch::2009}.
\newpage
\begin{description}
	\item[Gene selection table] Genes selected in the SUGSVarSel algorithm applied to the Golub dataset
\end{description}

\begin{figure}[H]\label{figure::golubgene1}
	\includegraphics[width=\linewidth, page=1]{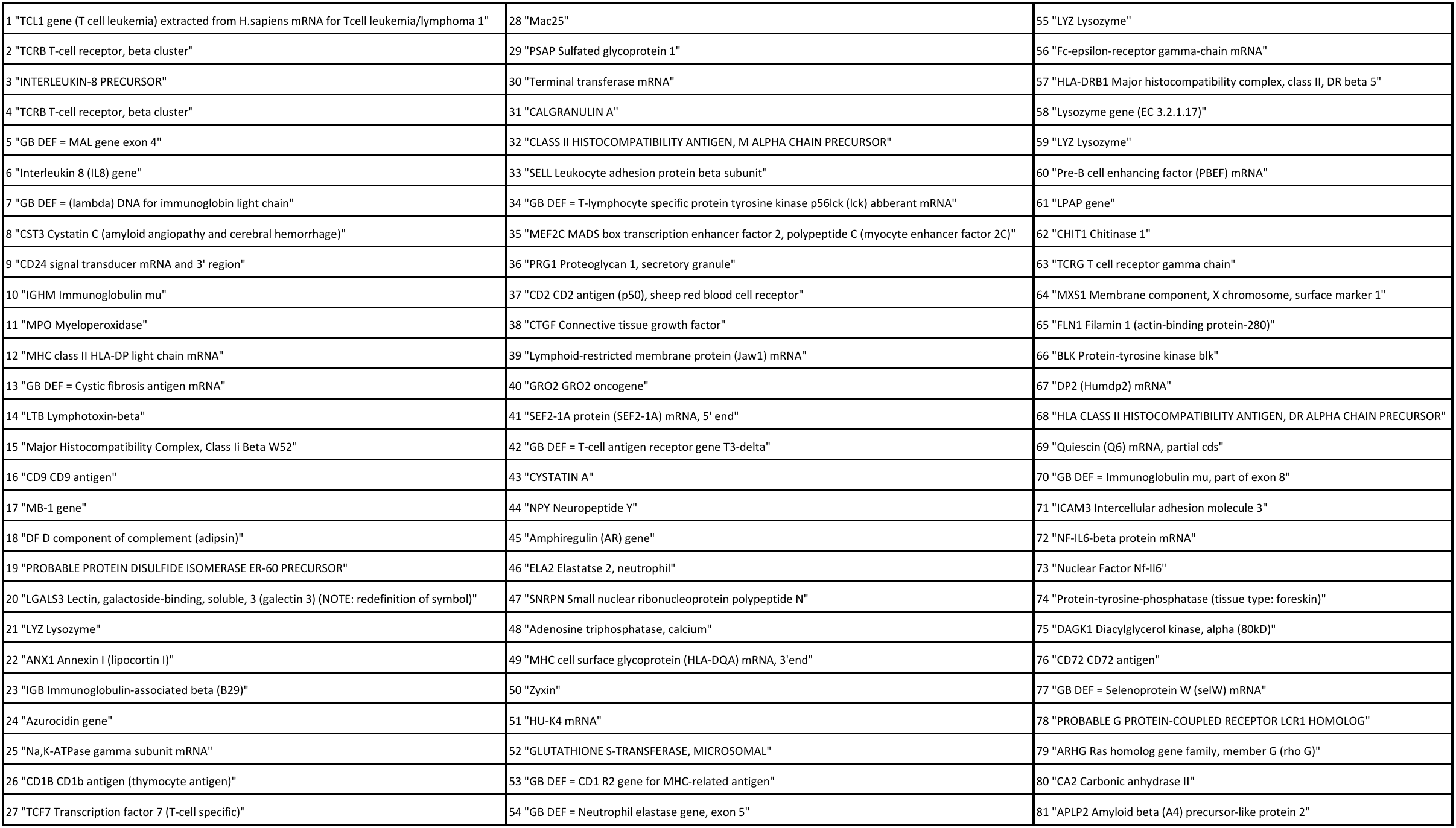}
	\centering
\end{figure}
\begin{figure}[H]\label{figure::golubgene2}
	\includegraphics[width=0.33\linewidth, page=2]{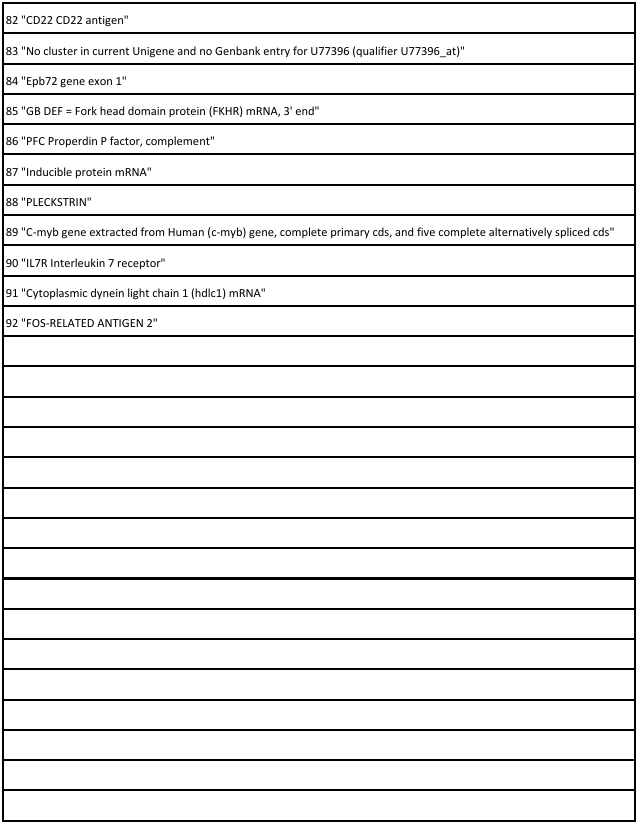}
	\centering
\end{figure}
\newpage

\begin{description}
\item[Pan-cancer proteomics cancer types]
\end{description}

The cancers used in the pan-cancer proteomic analysis are Uterine Corpus Endometrial Carcinoma (UCEC), Ovarian (OV), Kidney Renal Clear Cell Carcinoma (KIRC), Colon Adenocarcinoma (COAD), glioblastoma multiforme (GBM), Rectum adenocarcinoma (READ), Thyroid carcinoma (THCA), Kidney renal papillary cell carcinoma (KIRP), Pancreas adenocarcinoma (PAAD), Adenoid cystic carcinoma (ACC), Urothelial Bladder Carcinoma (BLCA), breast cancer (BRCA), Head and neck squamous cell carcinoma (HNSC), Lower grade glioma (LGG), Lung adenocarcinoma (LUAD), Lung Squamous Cell Carcinoma (LUSC), Prostate Adenocarcinoma (PRAD), skin cutaneous melanoma (metastatic and primary) (SKCM) and Stomach Adenocarcinoma (STAD).
\\

%\begin{description}
%	\item[Pan-cancer proteomics hierarchical clustering]
%\end{description}
%  In Figure~\ref{figure::heatTCPAhclust}, we visualise the data in a heatmap and perform hierarchical clustering on the tumour sample and variables using Euclidean distance \citep{Akbani::2014}.
%\begin{figure}[H]
%	\includegraphics[width=10cm]{TCPAheathclust.pdf}
%	\centering
%	\caption{A heatmap of the TCPA dataset with hierarchical clustering of tumours and variables. The dendrogram for the tumours is omitted for clarity, but the ordering is retained. The colour bar indicates the cancer types.}\label{figure::heatTCPAhclust}
%\end{figure}

\begin{description}
\item[Cancer clustering table] In Table~\ref{cancerClusters}, we summarise the correspondence between the 18 largest SUGSVarSel clusters (which omits clusters with fewer than 20 members) and the cancer types.  

\begin{table}[H]
	\caption{Number of tumours of each cancer type (rows) in each SUGSVarSel cluster (columns).}\label{cancerClusters}
	\centering
	\resizebox{\columnwidth}{!}{%
	\begin{tabular}{rrrrrrrrrrrrrrrrrrr}
		\hline
		& A & B & C & D & E & F & G & H & I & J & K & L & M & N & O & P & Q & R \\ 
		\hline
		UCEC & 325 & 43 & 1 & 1 & 11 & 2 & 0 & 0 & 0 & 0 & 0 & 0 & 0 & 0 & 0 & 0 & 0 & 0 \\ 
		OV & 363 & 42 & 0 & 0 & 0 & 2 & 0 & 0 & 0 & 0 & 0 & 0 & 0 & 0 & 0 & 0 & 0 & 0 \\ 
		KIRC & 3 & 25 & 0 & 2 & 0 & 0 & 405 & 0 & 0 & 0 & 0 & 0 & 0 & 0 & 0 & 0 & 0 & 0 \\ 
		COAD & 18 & 33 & 229 & 0 & 28 & 17 & 0 & 0 & 0 & 1 & 0 & 0 & 0 & 0 & 0 & 0 & 0 & 0 \\ 
		READ & 8 & 13 & 92 & 1 & 10 & 4 & 0 & 0 & 0 & 0 & 0 & 0 & 0 & 0 & 0 & 0 & 0 & 0 \\ 
		GBM & 5 & 6 & 0 & 4 & 0 & 0 & 0 & 170 & 2 & 0 & 0 & 0 & 0 & 0 & 0 & 0 & 0 & 0 \\ 
		THCA & 0 & 177 & 0 & 3 & 0 & 0 & 0 & 0 & 0 & 0 & 64 & 128 & 0 & 0 & 0 & 0 & 0 & 0 \\ 
		KIRP & 7 & 40 & 0 & 2 & 0 & 0 & 0 & 0 & 0 & 0 & 0 & 0 & 158 & 0 & 0 & 0 & 0 & 0 \\ 
		PAAD & 5 & 72 & 0 & 2 & 3 & 0 & 0 & 0 & 0 & 0 & 0 & 0 & 0 & 23 & 0 & 0 & 0 & 0 \\ 
		ACC & 6 & 6 & 0 & 0 & 1 & 0 & 0 & 0 & 0 & 0 & 0 & 0 & 33 & 0 & 0 & 0 & 0 & 0 \\ 
		BLCA & 66 & 46 & 1 & 0 & 3 & 0 & 0 & 0 & 0 & 6 & 0 & 0 & 0 & 0 & 0 & 0 & 0 & 0 \\ 
		BRCA & 151 & 192 & 0 & 34 & 0 & 0 & 0 & 0 & 0 & 1 & 0 & 0 & 0 & 0 & 415 & 0 & 0 & 0 \\ 
		HNSC & 24 & 49 & 0 & 1 & 16 & 3 & 0 & 0 & 0 & 104 & 0 & 0 & 0 & 0 & 0 & 1 & 0 & 0 \\ 
		LGG & 0 & 1 & 0 & 0 & 0 & 0 & 0 & 128 & 124 & 0 & 0 & 0 & 0 & 0 & 0 & 0 & 0 & 0 \\ 
		LUAD & 12 & 19 & 0 & 0 & 1 & 11 & 0 & 0 & 0 & 164 & 0 & 0 & 0 & 0 & 0 & 16 & 0 & 0 \\ 
		LUSC & 20 & 5 & 0 & 0 & 10 & 3 & 0 & 0 & 0 & 146 & 0 & 0 & 0 & 0 & 0 & 5 & 0 & 0 \\ 
		PRAD & 3 & 5 & 0 & 1 & 0 & 0 & 0 & 0 & 0 & 0 & 0 & 0 & 0 & 0 & 0 & 0 & 155 & 0 \\ 
		SKCM & 149 & 46 & 0 & 4 & 6 & 0 & 0 & 0 & 0 & 0 & 0 & 0 & 0 & 0 & 0 & 0 & 0 & 0 \\ 
		STAD & 8 & 66 & 99 & 2 & 25 & 45 & 0 & 0 & 0 & 0 & 0 & 0 & 0 & 0 & 0 & 0 & 0 & 38 \\ 
		\hline
	\end{tabular}%
}
\end{table}

\end{description}

%% BIBLIOGRAPHY
%\newpage
\bibliographystyle{apalike}
\bibliography{sugsReferences}

\end{document}